\documentclass[journal]{IEEEtran}
\usepackage{amsmath,amsfonts,mathtools,amsthm}
\usepackage{algorithmic}
\usepackage{algorithm}
\usepackage{array}
\usepackage[caption=false,font=normalsize,labelfont=sf,textfont=sf]{subfig}
\usepackage{textcomp}
\usepackage{stfloats}
\usepackage{url}
\usepackage{verbatim}
\usepackage{graphicx}
\usepackage{cite}
\usepackage{dsfont}
\usepackage[dvipsnames]{xcolor}
\usepackage{hyperref}
\usepackage{cleveref}

\definecolor{LightCoral}{RGB}{240, 128, 128}
\definecolor{NavajoWhite}{RGB}{255, 222, 173}
\definecolor{LightGoldenrodYellow}{RGB}{250, 250, 210}
\definecolor{LightGreen}{RGB}{144, 238, 144}

\newtheorem{theorem}{Theorem}

\newcommand{\new}[1]{{#1}}

\begin{document}

\title{Dynamic Dimensioning of Frequency Containment Reserves: The Case of the Nordic Grid}

\author{Jöbke Janssen, Alessandro Zocca,~\IEEEmembership{Member,~IEEE,}\\ Bert Zwart, Jalal Kazempour,~\IEEEmembership{Senior Member,~IEEE}%
\thanks{Jöbke Janssen and Bert Zwart are with the CWI, the national research institute of mathematics and computer science, in Amsterdam, The Netherlands; E-mail: jobke.janssen@cwi.nl, bert.zwart@cwi.nl}%
\thanks{Alessandro Zocca is with the VU, the Vrije Universiteit Amsterdam, in Amsterdam, The Netherlands; E-mail: a.zocca@vu.nl}%
\thanks{Jalal Kazempour is with the DTU, the Technical University of Denmark, in Copenhagen, Denmark; E-mail: jalal@dtu.dk}}%

\maketitle

\begin{abstract}
One of the main responsibilities of a Transmission System Operator (TSO) operating an electric grid is to maintain a designated frequency (e.g., 50 Hz in Europe). To achieve this, TSOs have created several products called frequency-supporting ancillary services. The Frequency Containment Reserve (FCR) is one of these ancillary service products. This article focuses on the TSO problem of determining the volume procured for FCR. Specifically, we investigate the potential benefits and impact on grid security when transitioning from a traditionally \textit{static} procurement method to a \textit{dynamic} strategy for FCR volume. We take the Nordic synchronous area in Europe as a case study and use a diffusion model to capture its frequency development. We introduce a controlled mean reversal parameter to assess changes in FCR obligations, in particular for the Nordic FCR-N ancillary service product. We establish closed-form expressions for exceedance probabilities and use historical frequency data as input to calibrate the model. We show that a dynamic dimensioning approach for FCR has the potential to significantly reduce the exceedance probabilities (up to $37\%$) while maintaining the total yearly procured FCR volume equal to that of the current static approach. Alternatively, a dynamic dimensioning approach could significantly increase security at limited extra cost.
\end{abstract}


\section{Introduction}
\label{Intro}
\IEEEPARstart{I}{n} modern electric grids, one of the critical responsibilities of Transmission System Operators (TSOs) is to maintain the grid’s target frequency---50 Hz in Europe and 60 Hz in the United States---to ensure system stability, balance real-time supply and demand, and support the functionality of connected devices. Achieving this stability demands precise interventions, for which TSOs have developed various frequency-supporting ancillary services tailored to address specific fluctuations and contingencies. 
These ancillary services vary between synchronous areas, as each grid has unique operational characteristics that shape the design of frequency-support mechanisms.

In the Continental Europe synchronous area, for example, TSOs employ three primary types of frequency containment and restoration services. Each service type has distinct response times, operating conditions, and purposes. The fastest is the Frequency Containment Reserve (FCR) service and is also called `primary reserve' or `regulating reserve'. FCR providers can bid in a market dedicated to the procurement of this service to provide power (i.e., capacity) for up or down regulation within 30 seconds, with the goal of \textit{containing} the frequency process. The second ancillary service product is the automatic Frequency Restoration Reserve (aFRR) and is also called `secondary reserve' or `operating reserve'. Here, providers should be able to provide a full response within 5 minutes, with the goal of \textit{restoring} the frequency back to 50 Hz. The third and `slowest' ancillary service product is the manual Frequency Restoration Reserve (mFRR) and is also called `tertiary reserve' or `replacement reserve'. Here, providers should provide a full response within 15-20 minutes, and the goal is also to \textit{restore} the frequency back to 50 Hz \cite{EnerginetPreQ}, \cite{EnerginetOutlook}.

Another key synchronous area in Europe is the Nordic synchronous area, which serves as the case study for this paper. This grid connects the electricity networks of Norway, Sweden, Finland, and the eastern part of Denmark. Due to its smaller size relative to the Continental Europe synchronous area (which spans 24 countries), the Nordic area experiences more pronounced frequency impacts from power disturbances, resulting from its inherently lower grid inertia~\cite{MEHIGAN,EntsoeInertia}. The Nordic TSOs employ additional ancillary services designed to provide rapid frequency response, particularly for primary reserves~\cite{EnerginetPreQ,EnerginetOutlook}.

\begin{figure}[t]
    \centering
    \includegraphics[width=\linewidth]{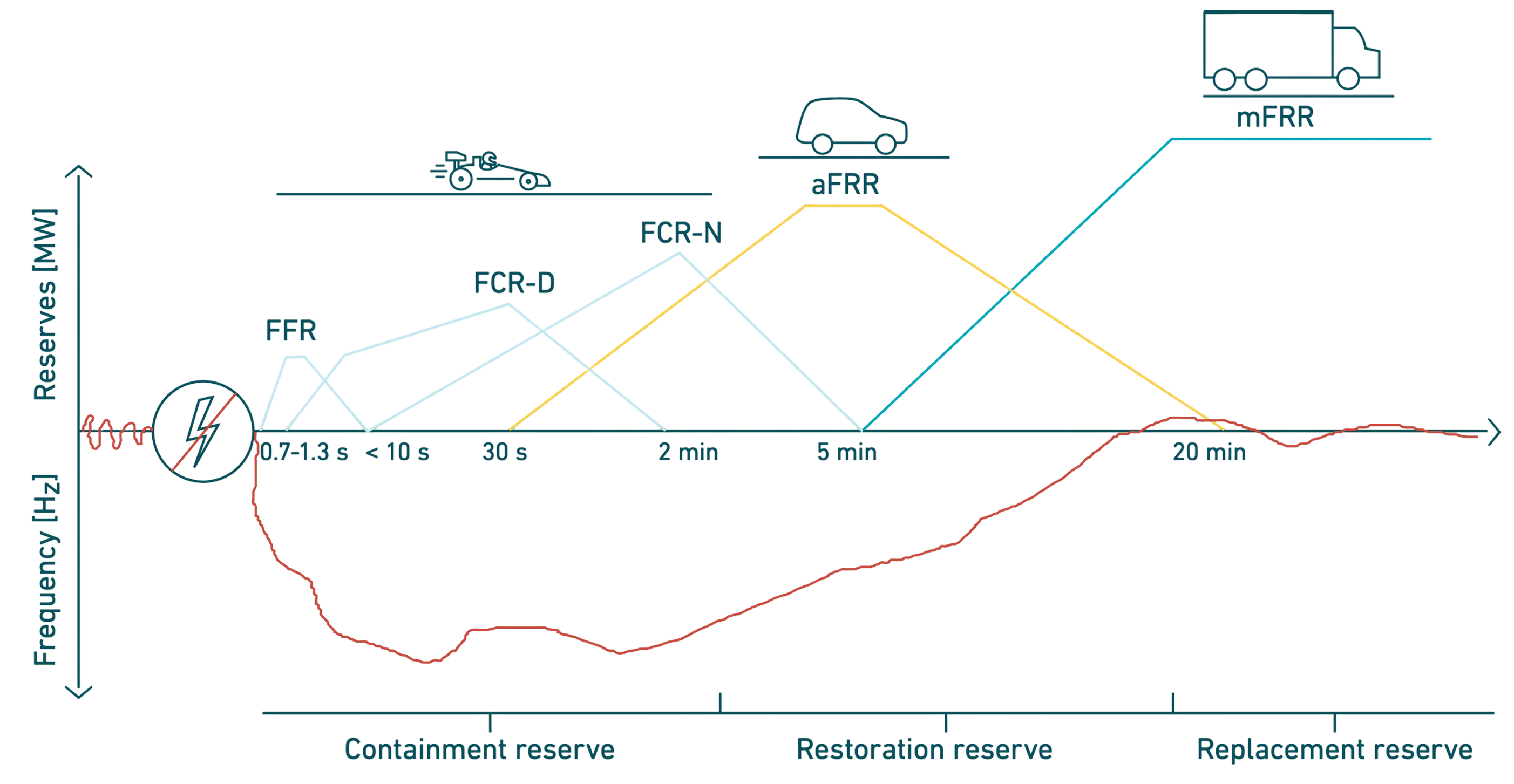}
    \caption{Activation times of ancillary services in Nordic synchronous area. Figure is borrowed from \cite{EnerginetOutlook}.}
    \label{fig:ASactivation}
\end{figure}

In particular, the Nordic TSOs utilize a specialized service called Fast Frequency Response (FFR) to immediately decelerate disturbances when grid frequency begins to deviate. In addition, the familiar Frequency Containment Reserve (FCR) service is used here but is adapted to the Nordic area’s needs: it differentiates between FCR-N for ``normal'' frequency deviations (49.9–50.1 Hz) and FCR-D for ``disturbed'' frequencies (below 49.9 Hz or above 50.1 Hz). Meanwhile, aFRR and mFRR reserves are also implemented in the Nordics, similar to their use in Continental Europe, with their alignment in the activation stage facilitated by the pan-European projects PICASSO (for aFRR) and MARI (for mFRR) \cite{EnerginetPreQ,EnerginetOutlook}. Figure \ref{fig:ASactivation} illustrates the activation timelines for these ancillary services within the Nordic area.

In the Nordics, the aFRR service is deployed proactively alongside FCR-N to address rising imbalances. However, with increased grid volatility due to electrification and a higher share of renewable generation, TSOs may soon face challenges maintaining frequency within safety margins \cite{Homan}. One possible strategy would be to increase the procured aFRR volumes. Although this could help moderate frequency volatility to some extent, the need for more responsive ancillary services, such as FCR, will grow as renewable penetration rising also yields a lower total inertia, thus shortening the time available to stabilize the grid during unexpected disturbances.

In this study, we examine the risk implications for grid security of adopting a more \textit{dynamic} approach to dimensioning the FCR-N volume, keeping FCR-D unchanged. Currently, the Nordic TSOs take a \textit{static} approach by procuring a constant volume of 600 MW for both up and down-regulation of FCR-N each hour throughout the year. This static volume of FCR-N is based on historical estimates of random load variations amounting to ±1\% of the 60 GW system load~\cite{Entsoe}. 

Our primary contribution is to challenge this static approach and investigate the impact of a dynamic FCR-N dimensioning method, wherein the volume procured varies hourly. 
Our findings indicate that this dynamic dimensioning strategy can reduce the likelihood of frequency deviations exceeding the $\pm 100$ mHZ band by as much as 37\% with respect to the current static approach, all while keeping the same total annual FCR-N volume procurement.

To provide these insights, we model frequency dynamics and the impact of FCR-N activation using a diffusion process framework. Recent studies have adopted similar stochastic process models to capture the behavior of various power system state variables, including line loads, line frequencies, and nodal voltages. For instance, historic frequency volatility in Great Britain is analyzed in \cite{Homan}, where a relationship is established between system inertia, frequency response delay, and ramp rates to ensure frequency requirements are met after a major infeed loss. A systematic modeling approach is introduced in \cite{AdeenMilano} using stochastic differential-algebraic equations to represent correlated stochastic processes in power systems, showing that correlations between processes can trigger instability.
In \cite{Deadbands}, the frequency probability density function is derived from a stochastic system model, revealing that system inertia has limited influence and that virtual inertia alone is insufficient to maintain nominal frequency under ambient load fluctuations. Reference \cite{Nesti} employs large deviation techniques to quantify the probability of current and temperature overloads in power grids with stochastic injections, and to determine corresponding safe capacity regions. Similarly, \cite{PatchZwart} identifies the most likely cause—conventional generator outages or renewable fluctuations—of frequency deviations reaching unacceptable thresholds as defined by the transmission system operator.
In \cite{Kruse}, a physics-informed machine learning model combines stochastic differential equations and neural networks to probabilistically model short-term frequency dynamics in the Continental European grid. A fast numerical approach is proposed in \cite{Sheng21} to efficiently compute frequency distributions in wind power systems, facilitating the analysis of how wind penetration and control parameters influence frequency behavior.
Among the studies above, \cite{Deadbands} and \cite{Sheng21} integrate diffusion-based models with deadband control to reproduce the empirically observed bimodal frequency distribution \cite{Kraljic}, and develop numerical procedures to solve the partial differential equations governing the frequency probability distribution function. In contrast, our approach models the FCR-N volume as a control variable within the drift term, leading to an analytically tractable model that captures the unimodal frequency distribution observed in the Nordic grid \cite{Kraljic}.

In addition, a related body of literature explores the dynamic dimensioning of ancillary services. However, these works focus on services such as aFRR and mFRR \cite{PoCBelgium, NSide, DDSweden}, whereas our study centers on FCR. While aFRR and mFRR are energy-intensive services, FCR typically is not, although it operates on a faster timescale. For aFRR and mFRR, factors such as system imbalances, contingencies, and available transmission capacity are crucial, which motivates the two-stage stochastic programming approach developed in \cite{DDSweden}. More broadly, probabilistic approaches to (static) reserve dimensioning have been applied in various contexts, including systems with high storage penetration \cite{Tsaousoglou}, high levels of wind integration \cite{N1security}, aggregator-based decision making \cite{Zhang}, and renewable forecast uncertainty using nonparametric methods and kernel-based estimators \cite{Bucksteeg}. To the best of our knowledge, however, no existing study specifically addresses the dynamic dimensioning of the FCR service.

The rest of the article is structured as follows. Section \ref{ProblemDescription} describes the ancillary services that support the 50 Hz grid frequency, with a focus on the specific requirements and dynamics of the Nordic synchronous area. Section \ref{Modelling} introduces the diffusion model used to capture frequency dynamics. In Subsection \ref{Definitions}, we present key definitions, and in Subsection \ref{Theory}, we derive closed form approximations for exceedance probabilities, quantifying the likelihood of frequency deviations within defined intervals such as between 49.5 and 49.9 Hz. In Section \ref{Data}, we describe the real world dataset containing over 300 million observations used to calibrate our model (Subsection \ref{D&M}) and present the main results through visual and quantitative analysis (Subsection \ref{Results}). Finally, Section \ref{Discussion} offers a summary and discussion, followed by future research directions in Section \ref{Outlook}. Proofs are included in the Appendix.

\section{Problem description} \label{ProblemDescription}
In this paper, we study how to \textit{dynamically} dimension the FCR ancillary service. In particular, we examine the potential benefits and impact on grid security when transitioning from the current \textit{static} procurement of FCR reserves to a \textit{dynamic} dimensioning approach, where the procured volume of FCR reserves depends on 
the volatility of the frequency, assuming the TSO can predict those\footnote{Forecasting the volatility parameter would require dedicated research, as any dynamic approach depends on predictive modeling. Similar efforts are already ongoing in industry. For example, the Belgian company N-SIDE currently collaborates with the Nordic TSOs and is already working on hourly day-ahead predictions of imbalances in the context of \textit{dynamic} dimensioning of aFRR/mFRR, based on the work they did for Elia, the Belgian TSO~\cite{NSide}.}. 
To assess the benefits of dynamic dimensioning, we compare the ``exceedance probabilities'', which quantify the likelihood that the frequency deviates beyond certain thresholds, e.g., $\pm 100$ mHz, from the target frequency of $50$ Hz at an arbitrary point in time. By examining these probabilities, we aim to determine how dynamic dimensioning affects grid stability in real-time. To support this analysis, in this section, we first provide a detailed overview of the relevant ancillary services, which will be captured in a mathematical model in the next section. We take the Nordic synchronous area as a case study; however, this approach can be adapted to other synchronous areas with potentially different ancillary services.

As mentioned in Section~\ref{Intro}, FCR in the Nordic synchronous area is differentiated into FCR-N for `normal' frequency conditions (i.e., frequencies between 49.9 and 50.1 Hz) and FCR-D for `disturbed' frequency conditions (i.e., frequencies below 49.9 Hz or above 50.1 Hz). Additionally, FCR-D is further segmented by up-regulation and down-regulation requirements, leading to three distinct FCR markets within the Nordic region.

FCR-N is sourced simultaneously for both up- and down-regulation, with a fixed obligation (i.e., to-be-procured volume) of 600 MW in both directions. This mechanism is designed to keep the exceedance probability outside of the normal band ($\pm 100$ mHz) to a maximum of 10,000 minutes per year ($\approx2\%$ of time)\cite{Entsoe}. FCR-N is activated proportionally to the frequency deviation from 50 Hz with a regulating strength of 6,000 MW/Hz between 49.9 and 50.1 Hz. Concretely, this means that 49.9 Hz corresponds to 600 MW of up-regulation and 50.1 Hz corresponds to 600 MW of down-regulation\cite{Entsoe2}.
Providers of FCR-N are paid for both reservation and activation. 

The obligation for FCR-D Up is fixed at 1,450 MW, which is equal to the largest reference incident for up-regulation in the Nordics.
FCR-D Up is activated proportionally to the frequency deviation from 50 Hz with a regulating strength of 3,625 MW/Hz between 49.5 and 49.9 Hz \cite{Entsoe2}.
The obligation for FCR-D Down is fixed at 1,400 MW, which is equal to the largest reference incident for down-regulation \cite{EnerginetOutlook}.
Similarly to FCR-D Up, FCR-D Down is activated proportionally to the frequency deviation between 50.1 and 50.5 Hz, but then in the other direction.

In the next section, where we model the FCR dynamics mathematically, we assume that FCR-D Up and FCR-D Down are sourced simultaneously in one market, similarly to FCR-N, ignoring the 50-MW difference for mathematical convenience. 

Finally, although our focus is on (dynamically) dimensioning FCR-N and establishing the probability of exceeding the $\pm 100$ mHz band, we also take into account the fixed obligation of FCR-D when modeling, since there is an obvious interplay between the two. Thus, we treat the fixed obligation of FCR-D as a (given) parameter in our model, while the FCR-N obligation is modeled as a (control) variable. A natural extension of this model is the one where the FCR-D obligation also becomes a control variable.

\section{Mathematical modeling} \label{Modelling}
We model the frequency development using a diffusion process that generalizes an Ornstein-Uhlenbeck process \cite{Klebaner}, \cite{BrowneWhitt} with a controlled mean reversal parameter that represents the linear activation nature of the FCR-N and FCR-D ancillary services as a function of the frequency deviation, resulting in a piecewise constant drift. In Subsection \ref{Definitions}, we define the necessary variables and introduce our model. In Subsection \ref{Theory}, we provide analytic expressions for the probability that the frequency is in a certain interval.

\subsection{Definitions and model}\label{Definitions}
We start with some preliminary definitions. Let
\begin{itemize}
    \item $F(t)$ be the shifted frequency in Hz (i.e., deviation from $50$ Hz, so that it has a zero mean) at time $t$;
    \item $r_N$ be the FCR-N obligation in GW, i.e., the to-be-procured volume of FCR-N reserve, for both up- and down-regulation. In this paper, $r_N$ is the control variable that can be optimized at each time interval (hourly, in our case);
    \item $x_D$ be the FCR-D obligation in GW, i.e., the to-be-procured volume of FCR-D reserve, for both up- and down-regulation, which is a (given) fixed parameter for all (hourly) time intervals;
    \item $\alpha(r_N;x_D, F(t)) \geq 0$ be the mean reversal function that describes the effect on the frequency due to activation of FCR-N and FCR-D ancillary services, see formula in \eqref{Alpha};
    \item $\mu(r_N;x_D,F(t),t):= -\alpha(r_N;x_D, F(t))F(t)$ be the drift function of the diffusion process;
    \item $\sigma(F(t),t):=\sigma \geq 0$ be the diffusion coefficient (`noise amplitude'), which is a fixed value for the time interval that is considered (hourly in this paper);
    \item $\{B(t)\}_{t\geq 0}$ be a zero-mean-unit-variance Brownian motion capturing the stochastic fluctuations.
\end{itemize}
The evolution over time of shifted frequency $F(t)$ can then be modeled by the following diffusion process:
\begin{align}
d{F}(t) &= \mu(r_N;x_D,F(t),t)dt + \sigma(F(t),t)d{B}(t) \nonumber\\
&= -\alpha(r_N;x_D, F(t))F(t)dt + \sigma d{B}(t), \label{Diffproc}
\end{align} 
where $\alpha(r_N, x_D; F)$ is defined as
\begin{align}
    &\alpha(r_N;x_D, F) = \nonumber \\
    \label{Alpha}&\begin{cases}
        r_N + x_D & \text{if } F < -0.5 \text{, or } F > 0.5\\
        r_N - \frac{1}{4}x_D - \frac{10}{4} x_D F  & \text{if } -0.5 \leq F \leq -0.1 \\
        -10 r_N F & \text{if } -0.1 \leq F \leq 0 \\
        10 r_N F & \text{if } 0 \leq F \leq 0.1  \\
        r_N - \frac{1}{4} x_D + \frac{10}{4} x_D F & \text{if } 0.1 \leq F \leq 0.5.
    \end{cases}
\end{align} 
\begin{figure}[t]
    \centering
    \includegraphics[width=\linewidth]{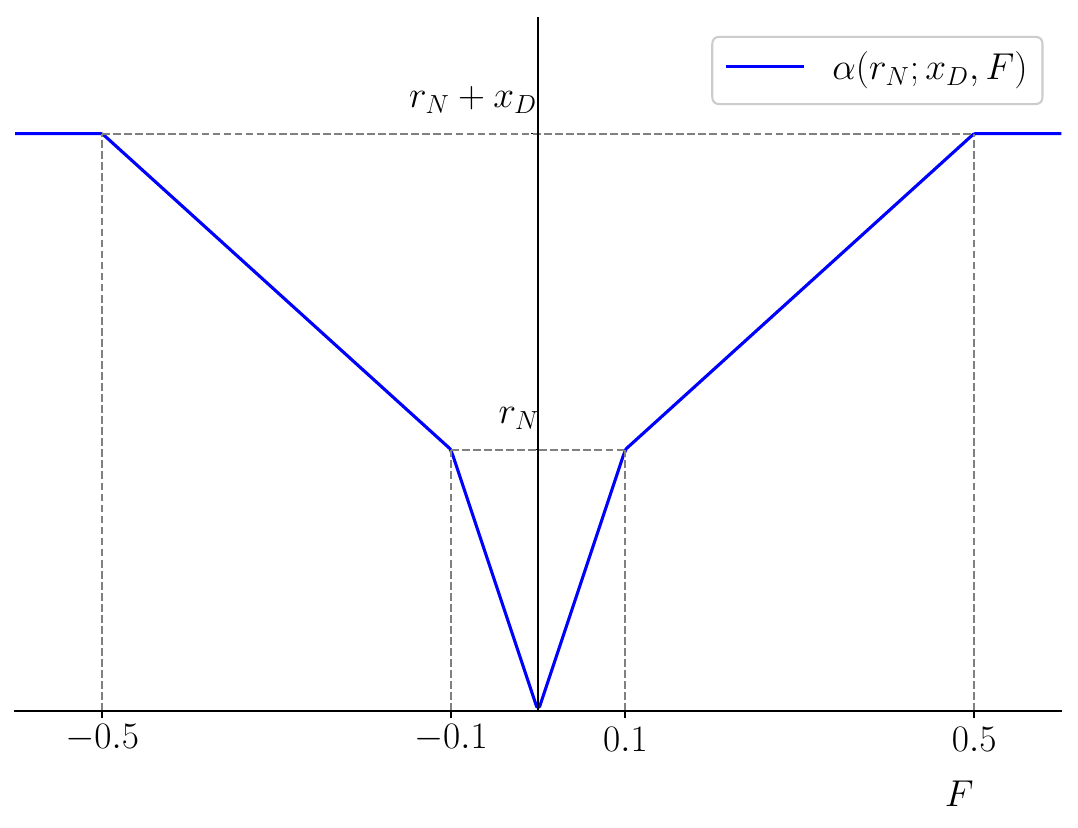}
    \caption{Graphic illustration of mean reversal function $\alpha(r_N;x_D, F)$, representing the activation of FCR-N and FCR-D according to formula \eqref{Alpha}.}
    \label{fig: alpha}
\end{figure}

\begin{figure}[t]
    \centering
    \includegraphics[width=\linewidth]{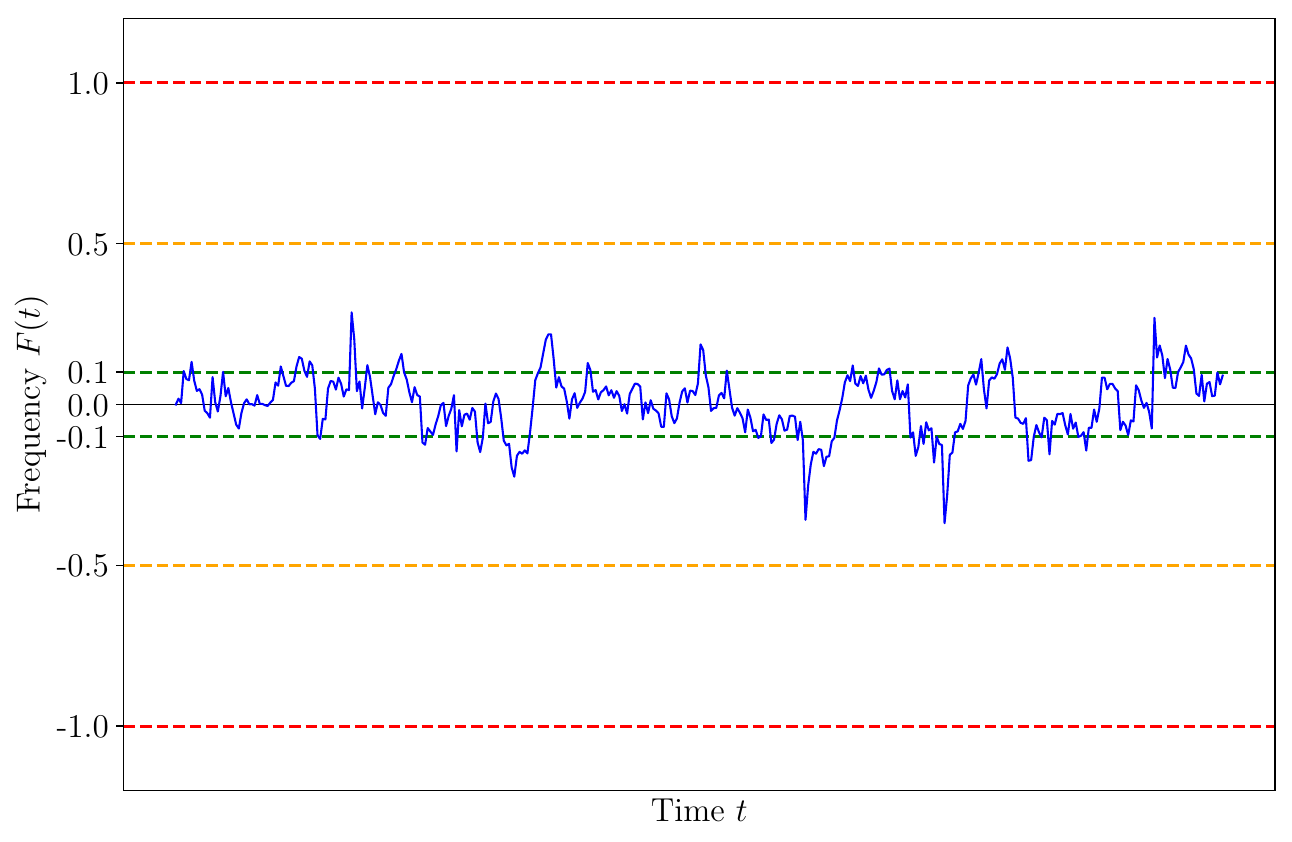}
    \caption{Random trajectory of the shifted frequency process $F(t)$ as prescribed by an Ornstein-Uhlenbeck process with state-dependent drift.}
    \label{fig:OU}
\end{figure}

Figure \ref{fig: alpha} illustrates the drift \eqref{Alpha} as a function of the frequency deviation. It can be easily checked that $\alpha(\cdot)$ is continuous in $F$. Since $-\alpha(\cdot; F) F <0$ if $F>0$ and $-\alpha(\cdot;F) F >0$ if $F<0$, it follows that the diffusion process \eqref{Diffproc} is indeed mean-reverting. Figure \ref{fig:OU} shows a simulation of the shifted frequency process $F(t)$ when using Ornstein-Uhlenbeck with the state-dependent drift as described by \eqref{Alpha}.

The main underlying assumptions of the above mathematical model are that activation of other ancillary services than FCR (i.e., FFR, aFRR, and mFRR) are included in the stochasticity of the Brownian motion. Also, we ignore network constraints and assume activation of FCR is instant, while, in reality, TSOs allow for ramp-up time. 

\subsection{Mathematical results} \label{Theory}
Using the mathematical model introduced in the previous subsection, we are now interested in determining exceedance probabilities to assess the benefits and risk implications of changing from a \textit{static} to a \textit{dynamic} way of dimensioning FCR-N. In other words, for a particular hour, given an estimate of the diffusion coefficient $\sigma$ (describing the current volatility of the frequency signal), a fixed volume $x_D$ for the FCR-D reserve, and a chosen volume $r_N$ for the FCR-N obligation, we aim to calculate the probability that $F(\cdot)$ is in a certain frequency interval. As a qualitative metric, we look at the fraction of time the frequency exceeds a given threshold, which is related to the invariant or stationary distribution of the frequency. We now provide analytic expressions for the aforementioned probability under the assumption that stationarity of the frequency is achieved within an hour, which is the time period for which we intend to make decisions.

To this end, in line with the function $\alpha(\cdot)$ as defined in \eqref{Alpha}, we define the relevant frequency intervals by 
\begin{align} \label{segments}
    \text{I}_1:= &  (-\infty, -0.5), \quad
    \text{I}_2:=  [-0.5,-0.1), \quad
    \text{I}_3:=  [-0.1,0), \nonumber\\
    \text{I}_4:= &  (0,0.1], \quad
    \text{I}_5:=  (0.1,0.5], \quad
    \text{I}_6:=  (0.5,+\infty).
\end{align}
Combining theory from \cite{Klebaner, BrowneWhitt}, we present the following result, which is proved in Appendix A.

\begin{theorem} \label{Thm: StatDistr}
Let $F(t)$ be the shifted frequency as defined in Section \ref{Modelling}. Its stationary distribution has a probability density function that can be expressed as
\begin{equation}
\tau(f) = \sum_{j=1}^6 p_j \tau_j(f) \mathds{1}(f\in \text{I}_j), \qquad f \in \mathbb{R}, \label{tau} 
\end{equation}
where for each $j=1,\dots,6$, $p_j>0$ and $\tau_j(f)$ are, respectively, the probability of being in interval $\text{I}_j$, and the probability density function in that interval, for which closed-form expressions are detailed below in \eqref{probab} and \eqref{taui}.
\end{theorem}

The six probabilities appearing in Theorem \ref{Thm: StatDistr}, that the shifted frequency $F$ is in specific intervals $\text{I}_j$, are given by
\begin{align}
p_1 & = p_6 = \frac{1}{2 + 2 \frac{K_2}{K_1} \text{exp}\left(-\frac{5x_D}{48\sigma^2}\right) + 2\frac{K_3}{K_1}\text{exp}\left(-\frac{31x_D+r_N}{300\sigma^2}\right)} \nonumber \\
p_2 & = p_5 = \frac{\frac{K_2}{K_1} \text{exp}\left(-\frac{5x_D}{48\sigma^2}\right)}{2 + 2 \frac{K_2}{K_1} \text{exp}\left(-\frac{5x_D}{48\sigma^2}\right) + 2\frac{K_3}{K_1}\text{exp}\left(-\frac{31x_D+r_N}{300\sigma^2}\right)} \label{probab}\\
p_3 & = p_4 = \frac{\frac{K_3}{K_1}\text{exp}\left(-\frac{31x_D+r_N}{300\sigma^2}\right)}{2 + 2 \frac{K_2}{K_1} \text{exp}\left(-\frac{5x_D}{48\sigma^2}\right) + 2\frac{K_3}{K_1}\text{exp}\left(-\frac{31x_D+r_N}{300\sigma^2}\right)}. \nonumber
\end{align}
and the component densities functions $\tau_j$ are given by
\begin{align}
    \tau_1(f) &=\tau_6(f) = \frac{\text{exp}\left(-f^2\frac{r_N+x_D}{\sigma^2}\right)}{K_1(r_N;x_D, \sigma)}, \text{ for } f \in \text{ I}_1 \text{ or I}_6\nonumber \\
    \tau_2(f) &= \tau_5(-f)=\frac{\text{exp}\left(-f^2\frac{4 \, r_N-x_D}{4 \, \sigma^2}+ f^3\frac{5 \, x_D}{3\sigma^2}\right)}{K_2(r_N; x_D, \sigma)},\label{taui} \\
    &\quad \quad \quad \quad \quad \quad \quad \quad \quad \quad \quad \quad \quad \text{ for } f \in \text{ I}_2 \text{ or I}_5 \nonumber \\
    \tau_3(f) &= \tau_4(-f) = \frac{\text{exp}\left(\frac{ 20 f^3 r_N}{3 \sigma^2}\right)}{K_3(r_N; \sigma)}, \quad \text{ for } f \in \text{ I}_3 \text{ or I}_4, \nonumber
\end{align}
with $K_1, K_2$ and $K_3$ normalizing constants defined as 
\begin{align}
K_1:=&K_1(r_N; x_D, \sigma):=\int_{-\infty}^{-0.5} \text{exp}\left(-y^2\frac{r_N+x_D}{\sigma^2}  \right)dy, \nonumber \\
K_2:=&K_2(r_N; x_D, \sigma):= \\
&\int_{-0.5}^{-0.1} \text{exp}\left(-y^2\frac{4 \, r_N-x_D}{4 \, \sigma^2} +y^3\frac{5 \, x_D}{3\sigma^2}\right)dy, \nonumber \\
K_3:=&K_3(r_N;\sigma):=\int_{-0.1}^{0.0} \text{exp}\left(\frac{20 y^3 r_N}{3 \sigma^2}\right)dy. \nonumber
\end{align}
Note that, by construction, $\tau$ is symmetric around $0$ and $\tau(f)$ is continuous since all the $\tau_j$'s are continuous and at the boundary points they coincide. 

In this paper we focus on the probability that the frequency deviations are large enough to exit from the $\pm 100$ mHZ band. Since the probability $p_1$ and $p_6$ of extreme variations turn out to be negligible (which is supported by historical data), in the rest of the paper we focus on $p_2$ (which is equal to $p_5$) as target exceedance probabilities.

\section{Nordics case study: Numerical results}\label{Data}
In this section, we look at real-world frequency data of the Nordic synchronous area from which we estimate the standard deviation $\sigma$ that serves as input for our mathematical model. In Subsection \ref{D&M}, we describe the data that we have used. Subsequently, in Subsection \ref{Results} we share our main findings. The Python code we used to obtain the results and create the figures is available at~\cite{Zocca2024}.

\subsection{Data}\label{D&M}
We obtained frequency data of the Nordic synchronous area from Finnish TSO, Fingrid \cite{Fingrid}, for the 12-month period from December 1, 2022, to November 30, 2023. The data has a time resolution of 0.1 seconds, resulting in over 300 million data points. We took 1-second averages of this raw data, in order to make computations less time-consuming, also aware that data points are rather correlated on the 0.1-second resolution. 
Alternative choices for the down-sampling (such as median, min, max, etc.) did not result in a significantly different volatility. We use these data to estimate $\sigma$ in \eqref{Diffproc} and set $x_D=1.45$ GW, as explained in Section \ref{ProblemDescription}.

\subsection{Results} \label{Results}
We now give an overview of our main findings, which relate to the benefits and risk effects in terms of grid security, of the current \textit{static} dimensioning approach versus the \textit{dynamic} dimensioning approach with respect to the FCR-N services. We show that under a \textit{static} dimensioning approach and with increasing volatility, one would need much more FCR-N volume than under a \textit{dynamic} dimensioning approach to meet the same safety levels. We introduce several heuristics for the dynamic dimensioning approach that focus on deploying significantly more FCR-N reserves in high-volatility periods and much less in low-volatility periods. We demonstrate that this dimensioning approach results in substantially more effective use of the FCR-N budget than the current static one.

\subsubsection{Static dimensioning}
\begin{figure}[t]
    \centering
    \includegraphics[width=\linewidth]{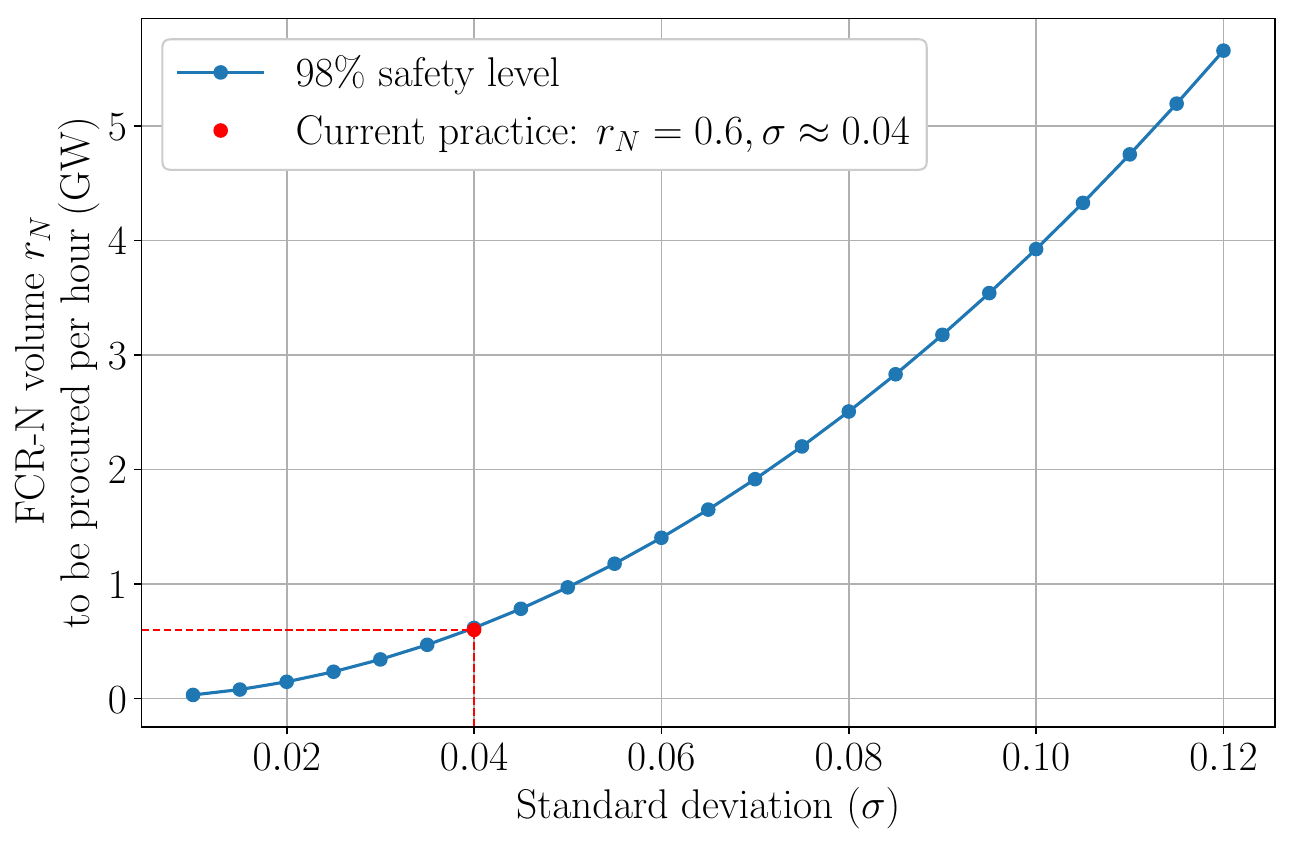}
    \caption{Increasing hourly FCR-N volumes $r_N$ (in GW), are required to make sure that $2p_2\leq 2\%$, (i.e., to stay within the +/- 100 mHz band $98\%$ of the time) for increasing $\sigma$.}
    \label{fig:FCR-N}
\end{figure}

We first study the current \textit{static} dimensioning approach to understand the risk effects in terms of grid security of increasing volatility on the required FCR-N volumes. Figure \ref{fig:FCR-N} shows, for varying $\sigma$, what FCR-N volumes ($r_N$) are required to stay within the $\pm 100$ mHz band for 98\% of the time, when the TSOs procure a constant FCR-N volume for every hour of the year. The 98\% \textit{safety level} stems from the Nordic TSOs' target to exceed the $\pm 100$ mHz for maximal 10,000 minutes per year. The Nordic TSOs currently procure a combined total of $r_N = 0.6$ GW of FCR-N volume for every hour of the year, where $\sigma \approx \, 0.04$. The main takeaway from Figure \ref{fig:FCR-N} is that, if volatility doubles (e.g., when $\sigma$ goes from ${\sim}0.04$ to ${\sim}0.08$) and no other measures are taken, then four times more FCR-N volume would be needed (i.e., instead of the current $r_N=0.6$ GW, TSOs would need to procure ${\sim}2.4$ GW per hour) to stay within the 98\% safety level under the current \textit{static} dimensioning approach.

\begin{figure}[t]
    \centering
    \includegraphics[width=\linewidth]{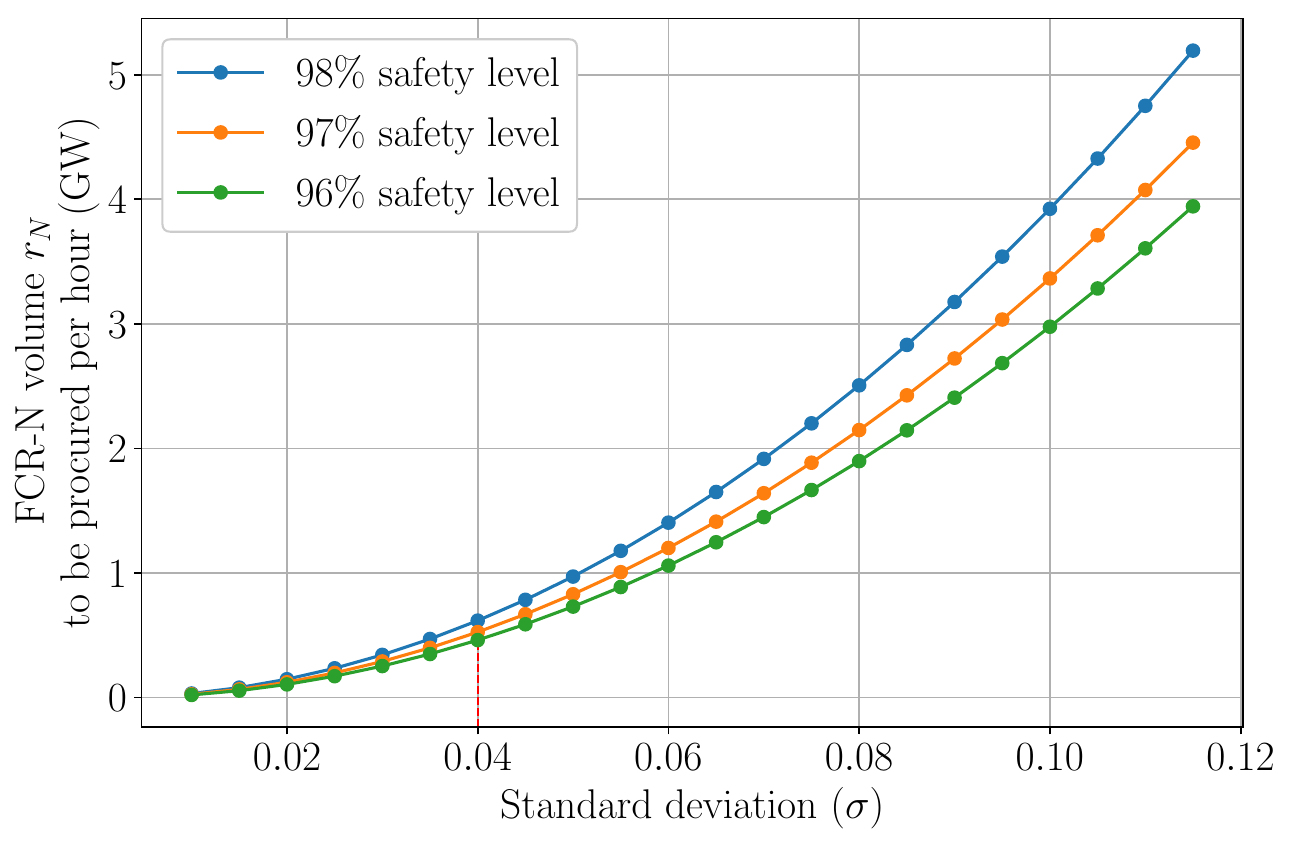}
    \caption{Required hourly FCR-N volumes $r_N$ (in GW) to stay within $98\%, 97\%$ and $96\%$ safety limits (i.e., required $r_N$ to make sure $2p_2 \leq 2\%$ or $2p_2 \leq 3\%$ or $2p_2 \leq 4\%$ holds, respectively) for varying $\sigma$.}
    \label{fig:SafetyLevels}
\end{figure}
Another option would be to lower the safety limits for the $\pm 100$ mHz band. This approach would automatically result in more frequent use of the FCR-D services, which might, in turn, result in a different required pricing mechanism for FCR-D since participants for FCR-D services currently only get paid for reservation (i.e., being stand-by), not for activation. Figure \ref{fig:SafetyLevels} shows the impact on the required FCR-N volumes $r_N$, when lower safety limits for the $\pm 100$ mHz band apply. As expected, if the Nordic TSOs decrease the safety limit, they need less FCR-N volumes to stay in the $\pm 100$ mHz band. However, doubling the risk tolerance (i.e., going from a safety level of $98\%$ to $96\%$) yields less than ${\sim}20\%$ lower required FCR-N volumes. Whether this is a sensible approach depends on the relative price differences between FCR-N and FCR-D. We do not go into this trade-off in this paper, also because, as mentioned above, a more frequent use of FCR-D might result in a different pricing mechanism, which is a research topic on its own.

\subsubsection{Dynamic dimensioning - a heuristic approach}
We propose to address the increasing volatility issue by applying a \textit{dynamic} dimensioning approach for the FCR-N service. This means that for each hour of the year, we determine a level of to-be-procured FCR-N volume, given expected day-ahead risk levels (i.e., expected volatility). We show that under `budget neutrality' (meaning the same total of annually procured FCR-N volume), higher safety levels can be achieved using the proposed \textit{dynamic} dimensioning approach instead of the current \textit{static} approach. In other words, keeping the 0.6 GW times 8,760 hours = 5,256 GW annual budget constant, we look at where it would be most effectively deployed.
There are, of course, multiple strategies for optimizing this allocation. In this paper, we adopt a heuristic approach that effectively illustrates our key findings while avoiding the complexities of computationally intensive optimization methods.

First, we define $r_N^i$, with $i=1,\ldots,8760$, as the amount of FCR-N volume reserved at hour $i$ in a year. 
Recall from \eqref{probab} that $p_2$ is the probability that the frequency is between 49.5 and 49.9 Hz (i.e., effectively, the probability that the frequency drops below 49.9 Hz, since $p_1$, the probability that the frequency drops below 49.5 Hz, is negligible, which is also supported by historical data). Defining $p_2^i$ as the probability that the frequency drops below 49.9 Hz in hour $i$, we propose a heuristic of the form
\begin{equation}
    r_N^i \quad = \quad \Bar{r}_N \, g(\sigma_i;\Bar{\sigma}) \, \eta \qquad \forall{i=1,\ldots,8760}, \label{DefrNi}
\end{equation}
where $\Bar{r}_N$ is the current constant level of procured FCR-N volume (i.e., $\Bar{r}_N=0.6$ GW), $\sigma_i$ is a day-ahead estimate of the expected standard deviation in hour $i$, $\Bar{\sigma}$ is the reference standard deviation (one could, e.g., take the average standard deviation of the previous year), $g(\cdot)$ is a to-be-defined function that scales the $\Bar{r}_N$ up or down depending on the expected day-ahead volatility and $\eta$ is a normalizing constant to make sure we end up within the total annual budget of 5,256 GW procured in a whole year. Note that in practice, determining $\eta$ upfront is impossible and hence one will not exactly end up with the exact annual budget for a full year (but for comparison purposes and since we look at historical data, we apply $\eta$ when comparing the various methods).

To ensure the heuristic follows the principle of scaling-up the FCR-N volume when high volatility is expected and vice versa, we impose the following properties on function $g(\cdot)$:
\begin{align} \label{Prop}
    0\leq g(\sigma_i;\Bar{\sigma}) \leq 1 & \text{ if } \sigma_i < \Bar{\sigma} \nonumber \\
    g(\sigma_i;\Bar{\sigma}) \geq 1 &\text{ if } \sigma_i > \Bar{\sigma} \\
    g(\sigma_i;\Bar{\sigma}) = 1 & \text{ if } \sigma_i = \Bar{\sigma} \nonumber 
\end{align}

Naturally, there are multiple sensible options to choose from with regard to $g(\cdot)$. One option is defining a threshold above/below which to deviate from $r_N^i=0.6$ GW. Other options would be to pick continuous functions, e.g., logarithmic, linear, or quadratic in $\sigma_i$. Since our primary goal of this paper is to show the potential benefits of employing a dynamic approach for FCR-N, we restrict ourselves to comparing a few simple options for $g(\cdot)$.

\subsubsection{Dynamic dimensioning heuristics with budget neutrality}
First, we consider a linear heuristic where $g(\cdot)=g^{\rm{Lin}}(\cdot)$, with
\begin{equation}
    g^{\rm{Lin}}(\sigma_i;\Bar{\sigma}) \quad := \quad \frac{\sigma_i}{\Bar{\sigma}},\label{glinear}
\end{equation}
which meets all properties in \eqref{Prop}. Then, 
\begin{equation}
    r_N^{\rm{Lin, i}} \quad \stackrel{\eqref{DefrNi}}{=} \quad \Bar{r}_N \, \frac{\sigma_i}{\Bar{\sigma}} \, \eta,  
\end{equation}
where $\eta$ is a parameter still to be determined. Since $g^{\rm{Lin}}$ is comparatively sensitive to the forecast standard deviation $\sigma_i$, it will probably be effective at lowering the overall exceedance probability. However, $r^{\rm{Lin,i}}_N$ can theoretically take on any non-negative value. Thus, we also consider a step-function heuristic $g(\cdot)=g^{\rm{Step}}(\cdot)$, with 
\begin{align}
    &g^{\rm{Step}}(\sigma_i;\Bar{\sigma}) \quad := \quad 
    \begin{cases}
        0.5 & \text{if } \sigma_i < \Bar{\sigma}\xi_{l} \\
        1 & \text{if } \Bar{\sigma} \xi_{l}  \leq \sigma_i \leq \Bar{\sigma} \xi^{u} \\
        1.5 & \text{if } \sigma_i > \Bar{\sigma} \xi^{u}, \label{gStep}
    \end{cases}
\end{align} 
with $\xi_{l}$ and $\xi^{u}$ as some lower and upper thresholds (e.g., $75\%$ and $125\%$), respectively. Then,
\begin{align}
    &r_N^{\rm{Step,i}} \quad \stackrel{\eqref{DefrNi}}{=} \quad 
    \begin{cases}
        0.5 \, \Bar{r}_N \, \eta & \text{if } \sigma_i < \Bar{\sigma}\xi_{l} \\
        \, \Bar{r}_N \, \eta & \text{if } \Bar{\sigma} \xi_{l}  \leq \sigma_i \leq \Bar{\sigma} \xi^{u} \\
        1.5 \, \Bar{r}_N \, \eta & \text{if } \sigma_i > \Bar{\sigma} \xi^{u}. 
   \end{cases}
\end{align} 
This step-function heuristic approach might be more acceptable for TSOs in practice than the linear one in~\eqref{glinear}, because the values are bounded and restricted to only three values, resulting in a simpler and more manageable market for FCR-N services.

\begin{figure}[t]
    \centering
    \includegraphics[width=\linewidth]{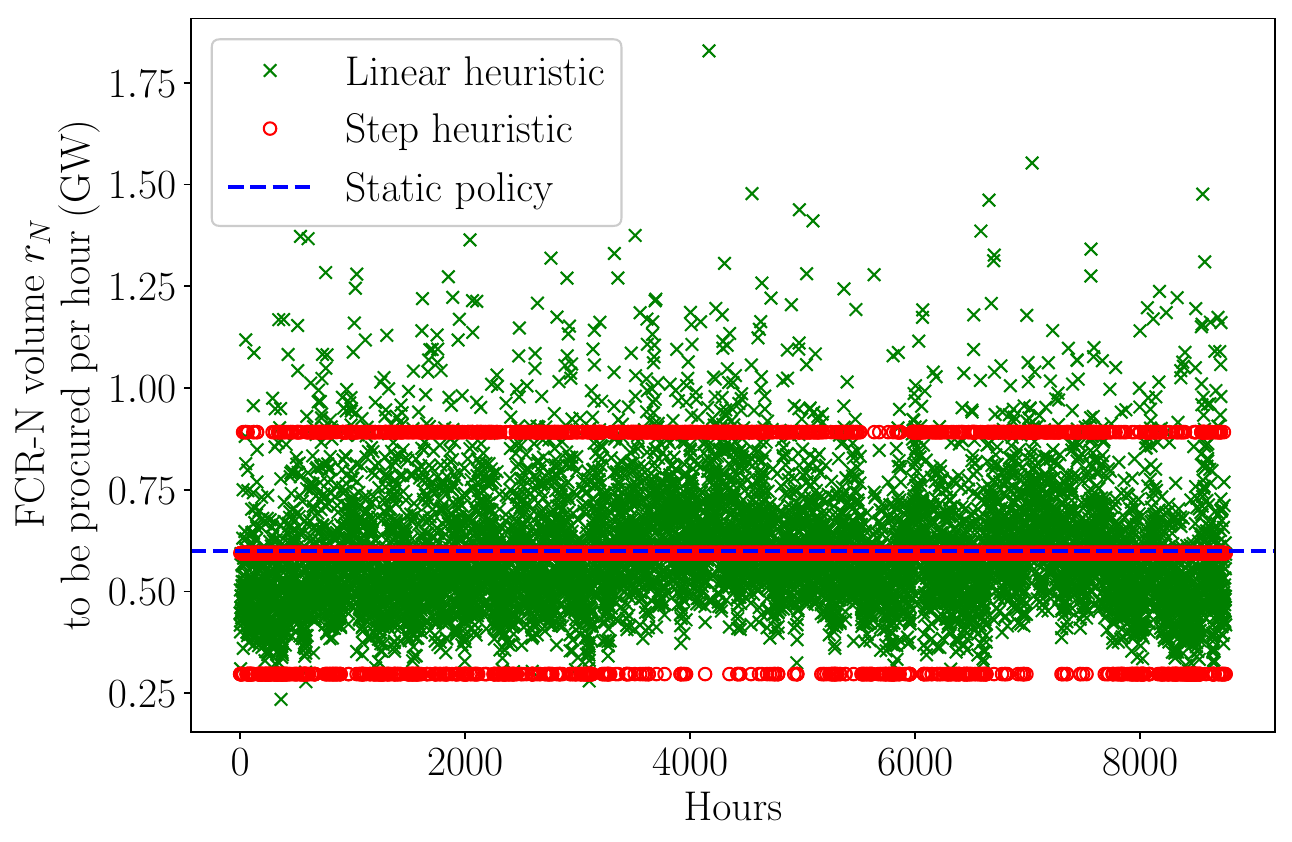}
    \caption{The distribution of $r_N^i$ volumes, for all $i=1,\ldots,8760$ hours in a year, is compared for the current \textit{static} approach and the two \textit{dynamic} approaches using the Linear and Step heuristics $g^{\rm{Lin}}$ and $g^{\rm{Step}}$.}
    \label{fig:heuristics-rN}
\end{figure}

Figure \ref{fig:heuristics-rN} shows the distribution of $r_N^i$ values for all 8,760 hours of the year for the \textit{dynamic} dimensioning approaches using $g^{\rm{Lin}}$ and $g^{\rm{Step}}$ compared to the flat line for the current \textit{static} approach. While all three approaches have the same annual budget of 5,256 GW of FCR-N volume (i.e., $\sum_{i=1}^{8760} r^i_N = 5,256$), the \textit{dynamic} approaches, by design, take on (many) different and more extreme $r_N^i$ values, as opposed to the \textit{static} one, where a constant level of 0.6 GW per hour is procured (i.e., $r_N^{\rm{Stat,i}}=0.6$ for all $i=1,\ldots,8760$).

\begin{figure}[t]
    \centering
    \includegraphics[width=\linewidth]{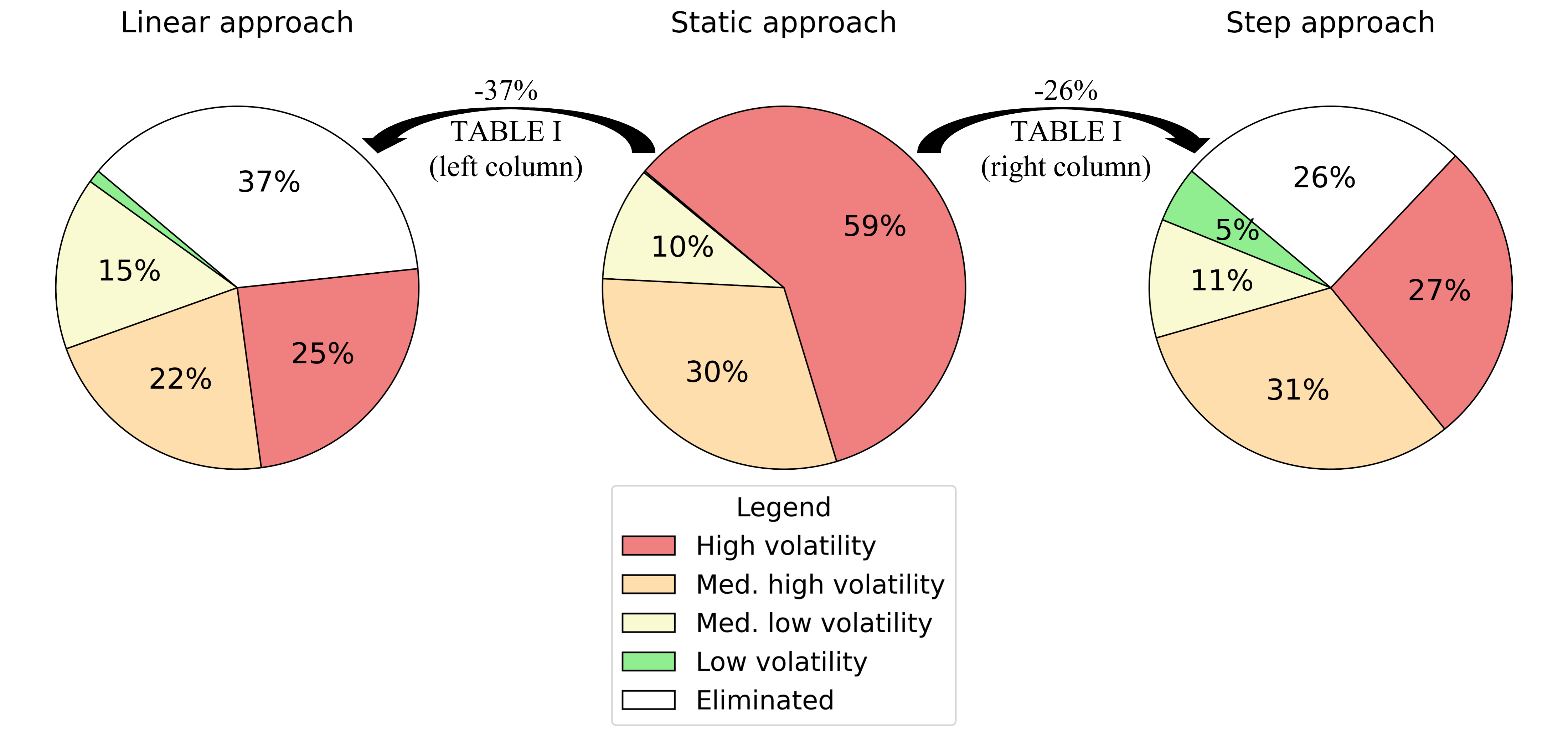}
    \caption{The relative size of exceedance probabilities $\left(\sum_i p_2^i\right)$ for different volatility segments is compared for the current \textit{static} approach and the two \textit{dynamic} approaches using the Linear and Step heuristics $g^{\rm{Lin}}$ and $g^{\rm{Step}}$.}
    \label{fig:heuristics-p2}
\end{figure}

The resulting annual exceedance probabilities (i.e., $\sum_{i=1}^{8760} p_2^i$, or the probability of getting out of the $\pm 100$ mHz band in a year) are the main contribution of this paper: The $g^{\rm{Lin}}$ and $g^{\rm{Step}}$ heuristics result in less extreme exceedance probabilities than the current \textit{static} approach as is illustrated in Figure \ref{fig:heuristics-p2} and Table \ref{table: lin and step}, where the volatility segments are defined as follows:
\begin{align}
    \text{`}\textbf{High volatility}\text{'} \text{ hours } \fcolorbox{black}{LightCoral}{\rule{0pt}{4pt}\rule{4pt}{0pt}}:& \, \text{All } i \text{ s.t. } \, \sigma_i>\xi_u \, \Bar{\sigma} \nonumber \\
    \text{`}\textbf{Medium high volatility}\text{'} \text{ hours } \fcolorbox{black}{NavajoWhite}{\rule{0pt}{4pt}\rule{4pt}{0pt}}:& \, \text{All } i \text{ s.t. } \, \Bar{\sigma} < \sigma_i \leq \xi_u \, \Bar{\sigma} \nonumber \\
    \text{`}\textbf{Medium low volatility}\text{'} \text{ hours } \fcolorbox{black}{LightGoldenrodYellow}{\rule{0pt}{4pt}\rule{4pt}{0pt}}:& \, \text{All } i \text{ s.t. } \, \xi_l \, \Bar{\sigma} < \sigma_i \leq \Bar{\sigma} \nonumber \\
    \text{`}\textbf{Low volatility}\text{'} \text{ hours } \fcolorbox{black}{LightGreen}{\rule{0pt}{4pt}\rule{4pt}{0pt}}:& \, \text{All } i \text{ s.t. } \, \sigma_i \leq \xi_l \, \Bar{\sigma},\label{volatilitysegments}
\end{align}
and the `\fcolorbox{black}{White}{\rule{0pt}{4pt}\rule{4pt}{0pt}} Eliminated' pie charts segments in Figure \ref{fig:heuristics-p2} indicate the overall decrease in exceedance probability of the $g^{\rm{Lin}}$ and $g^{\rm{Step}}$ approaches, when compared to the static one. 
In Figure \ref{fig:heuristics-p2}, the relative size of the pie charts (after subtracting the Eliminated parts) thus indicates the overall exceedance probability in a year ($\sum_{i=1}^{8760} p_2^i$), which is, as shown in the graph, highest for the \textit{static} approach. Observe that the exceedance probability in the \textit{static} case is driven mainly by the high volatility segment, as expected. Deploying extra FCR-N volume in times of high volatility and less in times of low volatility, as is the approach of the heuristics, results in a comparatively lower overall exceedance probability as highlighted by the size of the Eliminated parts of the pie charts (i.e., $37\%$ and $26\%$ for the $g^{\rm{Lin}}$ and $g^{\rm{Step}}$ heuristics, respectively). 
Table \ref{table: lin and step} underpins the drivers behind the above observations more quantitatively, where, for $i$ in one of the volatility segments as defined in \eqref{volatilitysegments}
\begin{align}
\Delta_{p_2}^{\rm{Lin}}&:=\frac{\sum_{i} p_2^{\rm{Lin,i}}-\sum_{i} p_2^{\rm{Stat,i}}}{\sum_{i} p_2^{\rm{Stat,i}}}, \text{ and} \label{Deltap2Lin}\\
\Delta_{r_N}^{\rm{Lin}}&:=\frac{\sum_{i} r_N^{\rm{Lin,i}}-\sum_{i} r_N^{\rm{Stat,i}}}{\sum_{i} r_N^{\rm{Stat,i}}}, \label{DeltarNLin}
\end{align}
are the relative change in exceedance probability and reserved FCR-N volume, respectively, for the $g^{\rm{Lin}}$ approach vs.~the \textit{static} (`Stat') one. $\Delta_{p_2}^{\rm{Step}}$ and $\Delta_{r_N}^{\rm{Step}}$ are defined similarly. 

Indeed, the heuristic approaches $g^{\rm{Lin}}$ and $g^{\rm{Step}}$ focus on deploying FCR-N services \textit{dynamically} mainly in times of high volatility ($+48\%$ and $+49\%$, respectively), resulting in lower overall exceedance probabilities (again, $-37\%$ and $-26\%$, respectively) compared to the current \textit{static} approach. Note that, relatively speaking, there is an enormous increase in exceedance probability in the low-volatile segment ($+654\%$ and $+3162\%$, respectively) for the $g^{\rm{Lin}}$ and $g^{\rm{Step}}$ approaches, but since these probabilities were very small to begin with, it has no significant impact on the overall exceedance probability (as noticeable in Figure \ref{fig:heuristics-p2}).

\begin{table}[t]
\centering
\renewcommand{\arraystretch}{1.5} 
\caption{Relative exceedance probabilities (left) \\ and fcr-n volumes (right) \\of $g^{\rm{Lin}}$ and $g^{\rm{Step}}$ vs. static approach}
\begin{tabular}{|l|r|r||r|r|}
\hline
\textbf{Volatility segment} & $\Delta_{p_2}^{\rm{Lin}}$ & $\Delta_{r_N}^{\rm{Lin}}$ & $\Delta_{p_2}^{\rm{Step}}$ & $\Delta_{r_N}^{\rm{Step}}$\\
\hline
\fcolorbox{black}{LightCoral}{\rule{0pt}{4pt}\rule{4pt}{0pt}} High volatility & $-59\%$ & $48\%$ & $-54\%$ & $49\%$\\
\fcolorbox{black}{LightGoldenrodYellow}{\rule{0pt}{4pt}\rule{4pt}{0pt}} Med. high volatility & $-29\%$ & $10\%$ & $3\%$ & $-1\%$\\
\fcolorbox{black}{NavajoWhite}{\rule{0pt}{4pt}\rule{4pt}{0pt}} Med. low volatility & $52\%$ & $-12\%$ & $4\%$ & $-1\%$\\
\fcolorbox{black}{LightGreen}{\rule{0pt}{4pt}\rule{4pt}{0pt}} Low volatility & $654\%$ & $-32\%$ & $3162\%$ & $-50\%$\\
\hline
\textbf{Overall} & $\mathbf{-37\%}$ & $\mathbf{0\%}$ & $\mathbf{-26\%}$ & $\mathbf{0\%}$ \\
\hline
\end{tabular}
\label{table: lin and step}
\end{table}

\subsubsection{Dynamic dimensioning heuristics with fixed safety levels}
In the above calculations, we assumed that the heuristics use the same annual budget of FCR-N services as the \textit{static} approach (namely, 5256 GW) and saw that this resulted in less extreme exceedance probabilities for the \textit{dynamic} dimensioning heuristics. Along the same lines, rather than equating the annual budget of FCR-N services, TSOs following a \textit{dynamic} approach could also choose to fix the exceedance probabilities at the same level as the \textit{static} approach. Consequently, TSOs could procure less FCR-N services while still obeying the same safety levels as for the \textit{static} approach. 

For example, let us define the heuristic $g^{\rm{ScaleLin}}$ which is equal to $g^{\rm{Lin}}$ up to a scale factor $\gamma$:
\begin{equation}
    g_{\gamma}^{\rm{ScaleLin}}(\sigma_i;\Bar{\sigma}):= g^{\rm{Lin}}(\sigma_i;\Bar{\sigma}) \gamma.
\end{equation}
Hence, the resulting $r^i_N$ is as follows:
\begin{equation}
    r^{\rm{ScaleLin,i}}_N:= g^{\rm{Lin}}(\sigma_i;\Bar{\sigma}) \gamma = \Bar{r}_N \, \frac{\sigma_i}{\Bar{\sigma}} \, \eta \, \gamma.
\end{equation}
By introducing $\gamma$ for the $g^{\rm{ScaleLin}}$ heuristic, we are scaling back $g^{\rm{Lin}}$ to an annual FCR-N budget that is just enough to meet the safety level of the \textit{static} approach.

\begin{figure}[t]
    \centering
    \includegraphics[width=\linewidth]{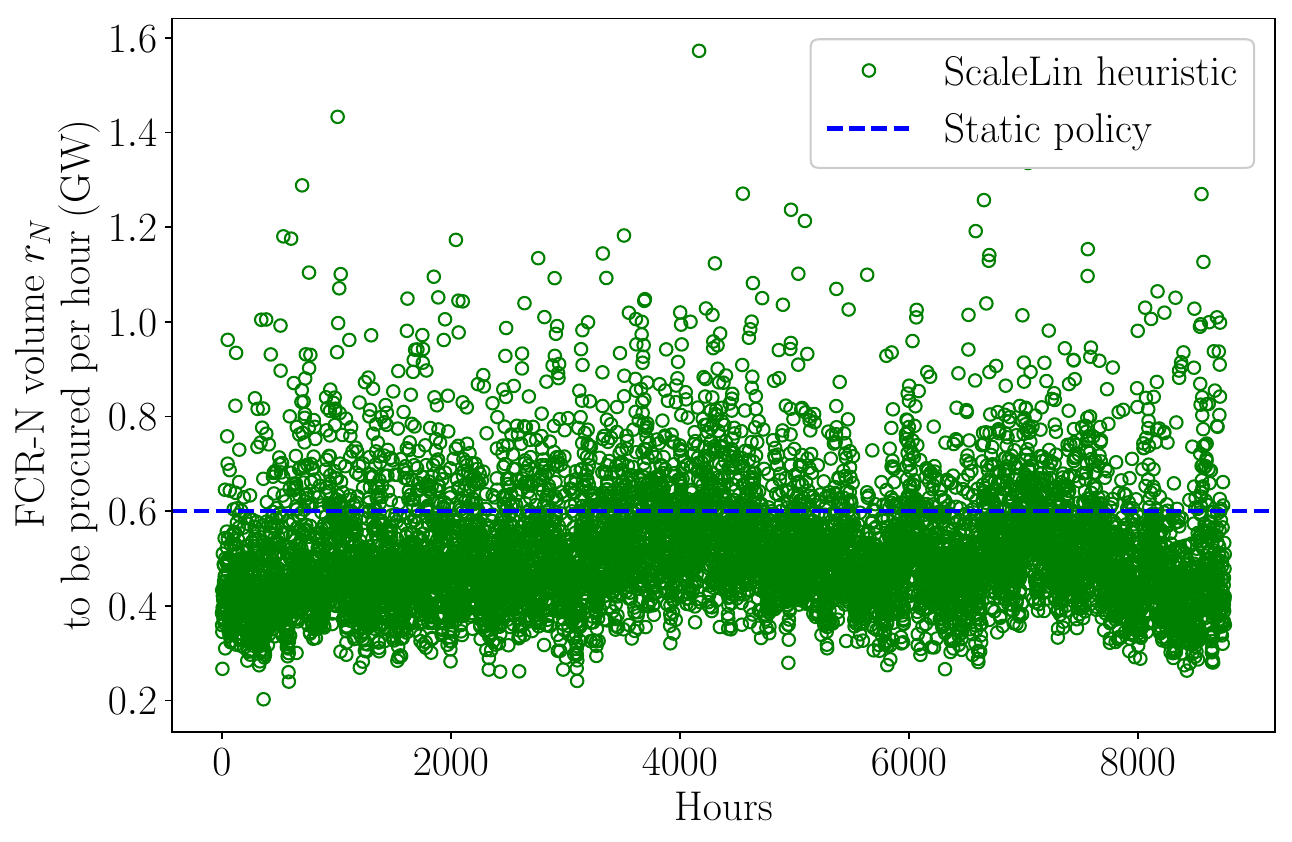}
    \caption{The distribution of $r_N^i$ volumes, for all $i=1,\ldots,8760$ hours in a year, is compared for the current \textit{static} approach and the \textit{dynamic} $g^{\rm{ScaleLin}}$ heuristic with $\gamma=0.86$.}
    \label{fig:ScaleLin-rN}
\end{figure}

Figure \ref{fig:ScaleLin-rN} shows the distribution of $r_N^i$ values for all 8,760 hours of the year for the \textit{dynamic} dimensioning approach using the $g^{\rm{ScaleLin}}$ heuristic (with $\gamma=0.86$) compared to the flat line for the current \textit{static} approach. Where the annual budget for the \textit{static} approach is 5,256 GW, the heuristic $g^{\rm{ScaleLin}}$ uses only ${\sim} 4,520$ GW of annual FCR-N budget, which is ${\sim}14\%$ less (or ${\sim}735$ GW), while still meeting the same safety level, as can also be concluded from Table \ref{table: scalelin}, where $\Delta_{p_2}^{\rm{ScaleLin}}$ and $\Delta_{r_N}^{\rm{ScaleLin}}$ are defined similarly as \eqref{Deltap2Lin} and \eqref{DeltarNLin}. 

\begin{figure}[t]
    \centering
    \includegraphics[width=\linewidth]{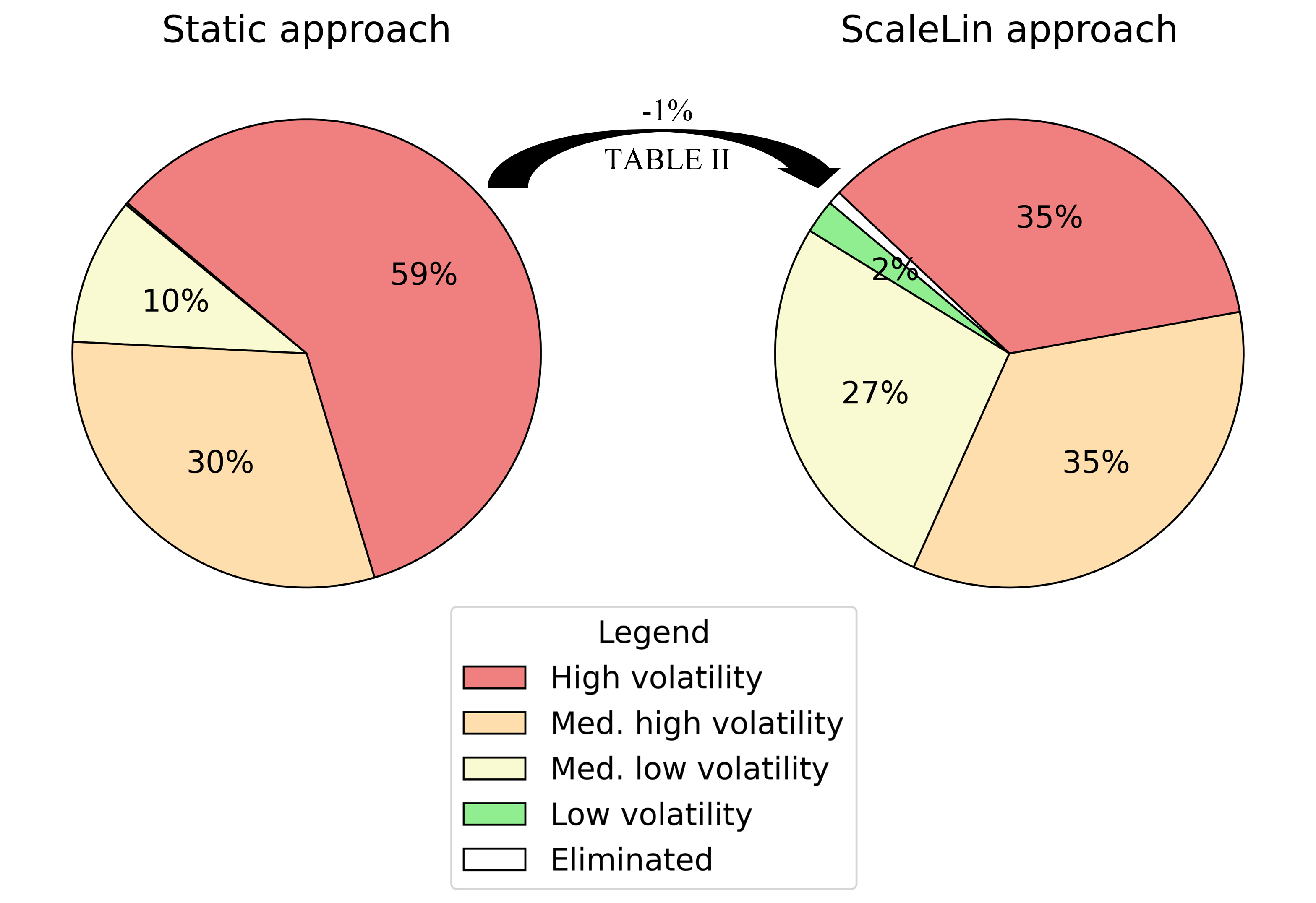}
    \caption{The relative size of exceedance probabilities $\left(\sum_i p_2^i\right)$ for different volatility segments is compared for the current \textit{static} approach and the \textit{dynamic} $g^{\rm{ScaleLin}}$ heuristic with $\gamma=0.86$.}
    \label{fig:ScaleLin-p2}
\end{figure}

In Figure \ref{fig:ScaleLin-p2}, the relative size of the pie charts (after subtracting the eliminated parts) again indicates the overall exceedance probability in a year ($\sum_{i=1}^{8760} p_2^i$), which is, as can be observed in the graph, approximately equal for both approaches. However, in the \textit{static} approach, the overall exceedance probability is mainly driven by hours with high volatility, while for the \textit{dynamic} approach using the $g^{\rm{ScaleLin}}$ heuristic, all segments except the low-volatility segment contribute significantly. Table \ref{table: scalelin} shows the drivers behind this more quantitatively. Indeed, the heuristic $g^{\rm{ScaleLin}}$ has approximately the same expected probability of exceedance over a full year (even ${\sim}1\%$ lower), but the distribution for different levels of volatility is completely different when compared to the \textit{static} approach. This difference is driven by procuring ${\sim}27\%$ more FCR-N services in high-volatility hours while buying less in lower-volatility hours, resulting in less overall required FCR-N volumes ($-14\%$) to achieve a similar level of safety.

\begin{table}[t]
\centering
\renewcommand{\arraystretch}{1.5} 
\caption{Relative exceedance probabilities (left) and \\ fcr-n volumes (right) of $g_{0.86}^{\rm{ScaleLin}}$ vs. static approach}
\begin{tabular}{|l|r|r|}
\hline
\textbf{Volatility segment} & $\Delta_{p_2}^{\rm{ScaleLin}}$ & $\Delta_{r_N}^{\rm{ScaleLin}}$ \\
\hline
\fcolorbox{black}{LightCoral}{\rule{0pt}{4pt}\rule{4pt}{0pt}} High volatility & $-41\%$ & $27\%$ \\
\fcolorbox{black}{LightGoldenrodYellow}{\rule{0pt}{4pt}\rule{4pt}{0pt}} Med. high volatility & $13\%$ & $-5\%$ \\
\fcolorbox{black}{NavajoWhite}{\rule{0pt}{4pt}\rule{4pt}{0pt}} Med. low volatility & $167\%$ & $-24\%$ \\
\fcolorbox{black}{LightGreen}{\rule{0pt}{4pt}\rule{4pt}{0pt}} Low volatility & $1423\%$ & $-41\%$ \\
\hline
\textbf{Overall} & $\mathbf{-1\%}$ & $\mathbf{-14\%}$ \\
\hline
\end{tabular}
\label{table: scalelin}
\end{table}

\subsubsection{Conservative dynamic dimensioning heuristic}
\begin{figure}[t]
    \centering
    \includegraphics[width=\linewidth]{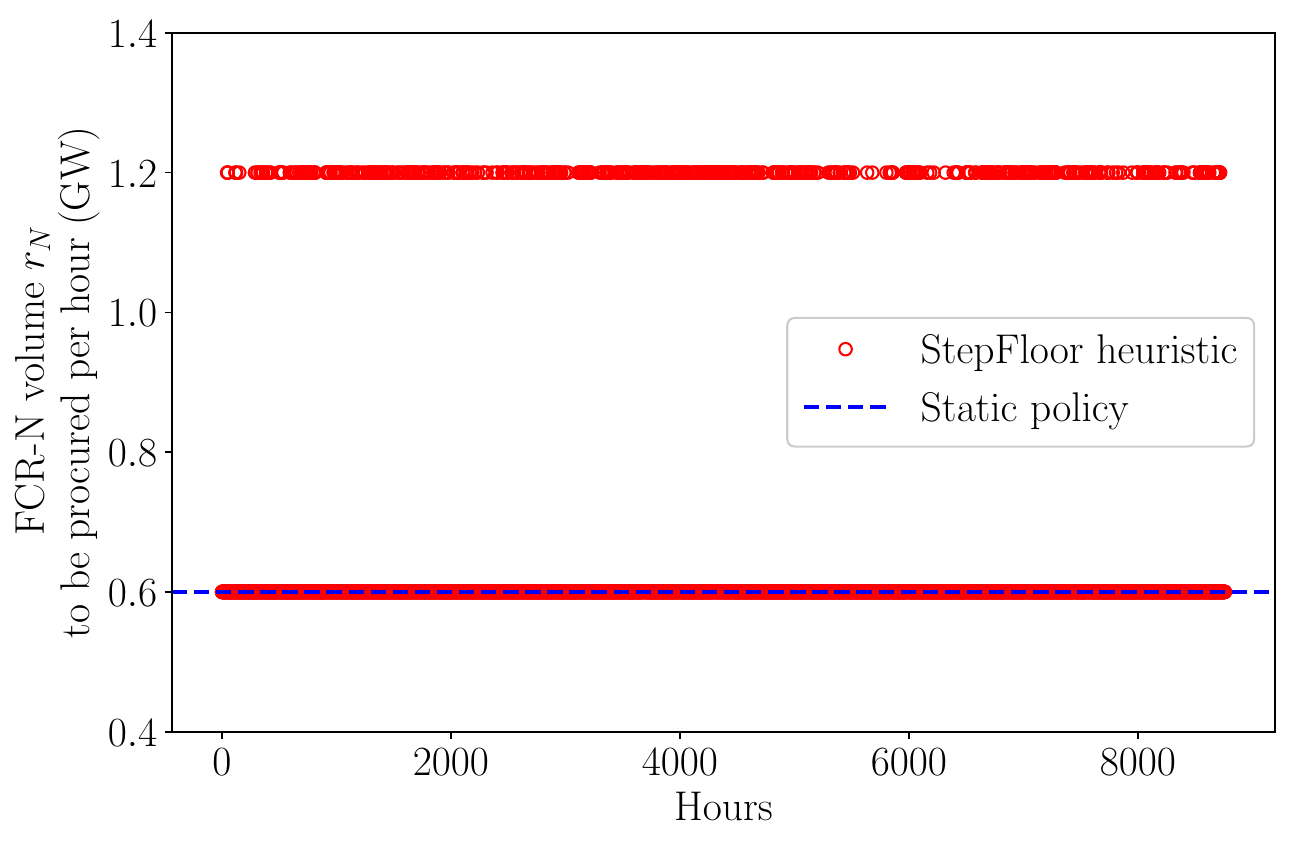}
    \caption{The distribution of $r_N^i$ volumes, for all $i=1,\ldots,8760$ hours in a year, is compared for the  current \textit{static} approach and the \textit{dynamic} $g^{\rm{StepFloor}}$ heuristic for $m=2$ and $\zeta=1.35$.}
    \label{fig:StepFloor-rN}
\end{figure}

Another, more conservative, approach to the dimensioning question, is when TSOs are hesitant to abandon the historical volume of $\Bar{r}_N = 0.6$ GW as minimum FCR-N obligation. We show that \textit{extra} FCR-N budget allocated at the right moments could reduce the risk of exceeding the $\pm 100$ mHz band significantly. In particular, one could selectively choose to procure more in high-volatility periods. For instance, one could alter $g^{\rm{Step}}$ by introducing the $\Bar{r}_N$ as minimum (`floor'), which would result in
\begin{align}
    &g^{\rm{StepFloor}}(\sigma_i;\Bar{\sigma}) \quad = \quad 
    \begin{cases}
       1 & \text{if } \sigma_i \leq \Bar{\sigma} \zeta \\
        m & \text{if } \sigma_i > \Bar{\sigma} \zeta, 
    \end{cases} \label{StepFloor}
\end{align} 
with $m$ a multiple (e.g., 2) of the constant $\Bar{r}_N$ and $\zeta$ some threshold level (e.g., 135 \%), respectively. Then, with $\eta=1$, we get
\begin{align}
    &r_N^{StepFloor, i} \quad \stackrel{\eqref{DefrNi}}{=} \quad 
    \begin{cases}
        \Bar{r}_N & \text{if } \sigma_i \leq \Bar{\sigma} \zeta \\
        m \, \Bar{r}_N & \text{if } \sigma_i > \Bar{\sigma} \zeta. 
    \end{cases}
\end{align} 
Figure \ref{fig:StepFloor-rN} shows the distribution of $r_N^i$ values for all 8,760 hours of the year for the \textit{dynamic} dimensioning approach using the $g^{\rm{StepFloor}}$ heuristic (with $m=2$ and $\zeta=1.35$) compared to the current \textit{static} approach. Where the annual budget for the \textit{static} approach is 5,256 GW, the heuristic $g^{\rm{StepFloor}}$ uses ${\sim} 5,700$ GW of annual FCR-N budget in this case, which is ${\sim}8\%$ (or ${\sim}445$ GW) more, as can also be observed in Table \ref{table: stepfloor}, where $\Delta_{p_2}^{\rm{StepFloor}}$ and $\Delta_{r_N}^{\rm{StepFloor}}$ are defined similarly as \eqref{Deltap2Lin} and \eqref{DeltarNLin}.

\begin{figure}[t]
    \centering
    \includegraphics[width=\linewidth]{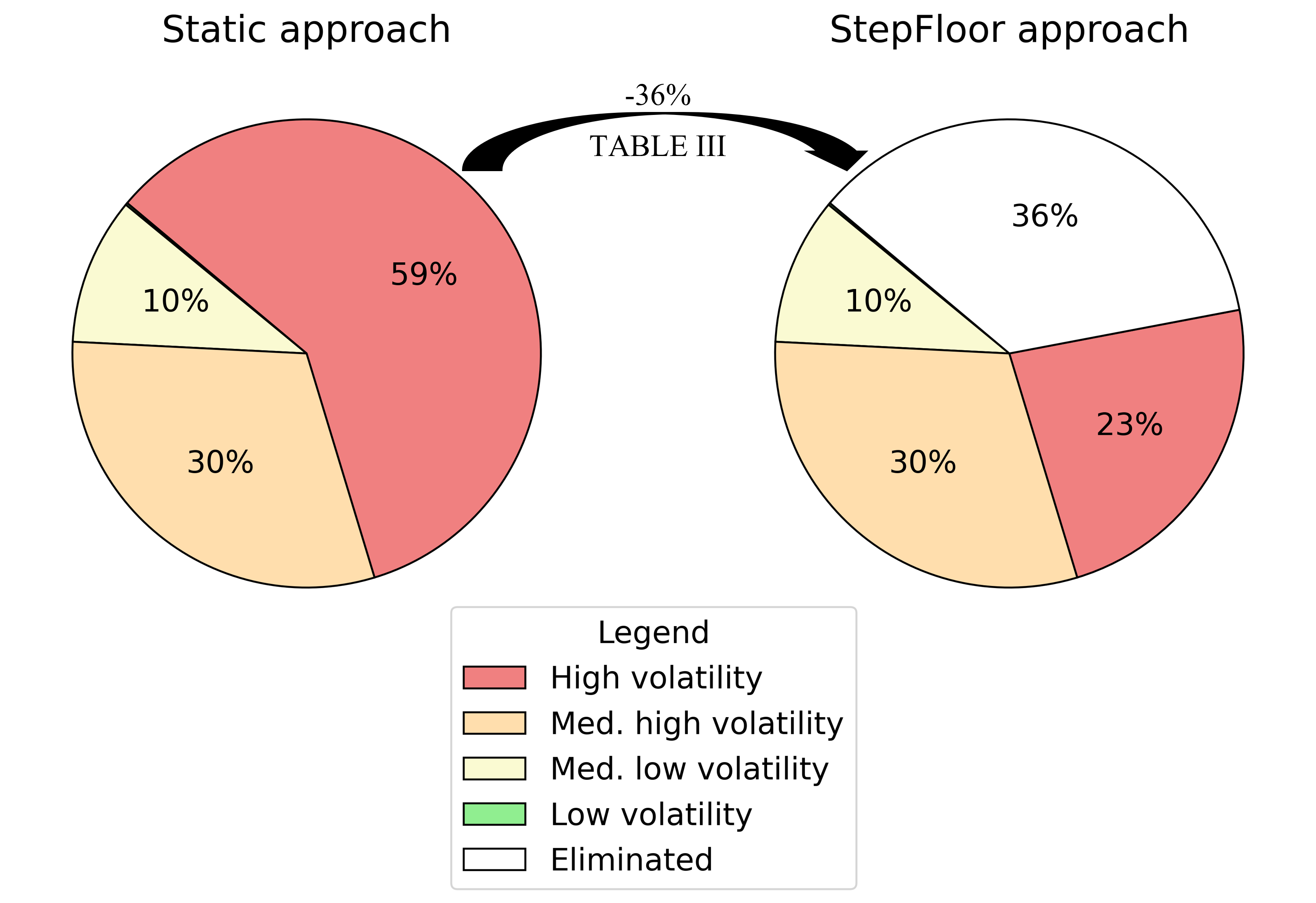}
    \caption{The relative size of exceedance probabilities $\left(\sum_i p_2^i\right)$ for different volatility segments is compared for the current \textit{static} approach and the \textit{dynamic}  
    $g^{\rm{StepFloor}}$ heuristic for $m=2$ and $\zeta=1.35$.}
    \label{fig:StepFloor-p2}
\end{figure}

In Figure \ref{fig:StepFloor-p2}, the relative size of the pie charts (after subtracting the eliminated part) indicates the overall exceedance probability in a year ($\sum_{i=1}^{8760} p_2^i$), which is, as expected, lower for the \textit{dynamic} $g^{\rm{StepFloor}}$ heuristic than for the \textit{static} approach. Indeed, Table \ref{table: stepfloor} shows that the extra ${\sim}8\%$ of annual FCR-N budget that $g^{\rm{StepFloor}}$ uses is fully deployed in the high-volatility segment ($+64\%$), resulting in a much lower overall exceedance probability ($-36\%$) driven by a strong reduction in exceedance probability in the high-volatility segment ($-61\%$).

\begin{table}[t]
\centering
\renewcommand{\arraystretch}{1.5} 
\caption{Relative exceedance probabilities (left) and \\ fcr-n volumes (right) of $g^{\rm{StepFloor}}$ vs. static approach}
\begin{tabular}{|l|r|r|}
\hline
\textbf{Volatility segment} & $\Delta_{p_2}^{\rm{StepFloor}}$ & $\Delta_{r_N}^{\rm{StepFloor}}$ \\
\hline
\fcolorbox{black}{LightCoral}{\rule{0pt}{4pt}\rule{4pt}{0pt}} High volatility & $-61\%$ & $64\%$ \\
\fcolorbox{black}{LightGoldenrodYellow}{\rule{0pt}{4pt}\rule{4pt}{0pt}} Med. high volatility & $0\%$ & $0\%$ \\
\fcolorbox{black}{NavajoWhite}{\rule{0pt}{4pt}\rule{4pt}{0pt}} Med. low volatility & $0\%$ & $0\%$ \\
\fcolorbox{black}{LightGreen}{\rule{0pt}{4pt}\rule{4pt}{0pt}} Low volatility & $0\% $& $0\%$ \\
\hline
\textbf{Overall} & $\mathbf{-36\%}$ & $\mathbf{8\%}$ \\
\hline
\end{tabular}
\label{table: stepfloor}
\end{table}

\subsubsection{Sensitivity to forecasting error} \label{sensitivity}

\begin{figure}[t]
    \centering
    \includegraphics[width=\linewidth]{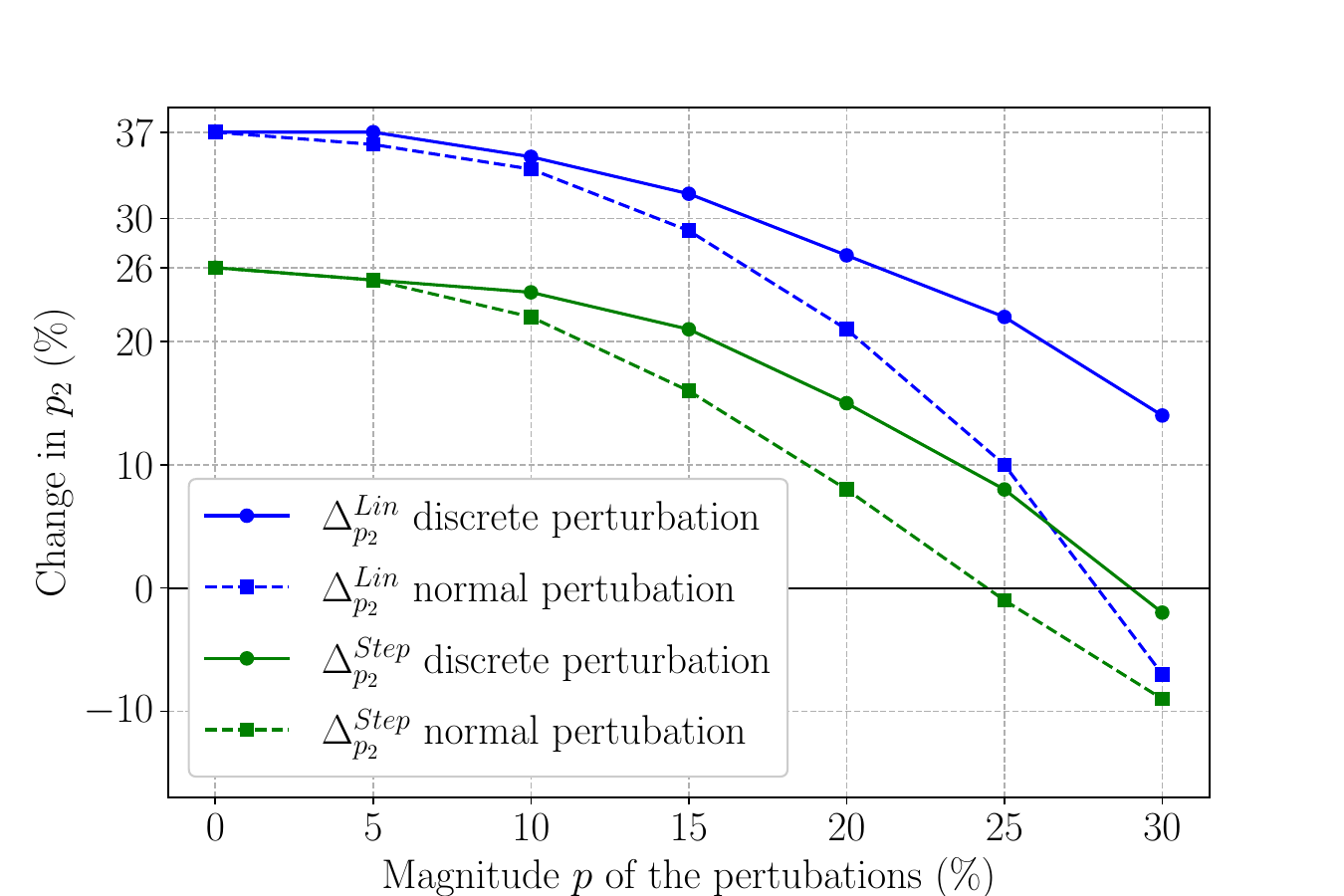}
    \caption{The performance of the dynamic $g^{Linear}$ and $g^{Step}$ heuristics, compared to the static approach in the presence of forecast errors of the $\sigma_i$ of increasing magnitude.}
    \label{fig:perturbations}
\end{figure}

The heuristics outlined in the previous subsections rely on the hourly standard deviations $\sigma_i$'s, which in an operational setting need to be estimated (e.g., the day ahead) and are thus subject to forecasting errors. In this subsection, we investigate the robustness of the proposed heuristics (Linear and Step) against perturbations of the hourly standard deviations. More specifically, we artificially introduce noise, obtaining a new time series $\{\hat \sigma_i\}_{i=1,\dots,8760}$ from the original one as $\hat\sigma_i = \psi_i \sigma_i$, where the $\psi=\psi_i(p)$'s are i.i.d.~random factors parametrized by a scalar $p \geq 0$ describing the magnitude of the perturbation.
We consider two different types of randomness for the factors $\psi_i(p)$'s. Specifically, we sample the random factors $\psi_i(p)$, $i=1,\dots,8760$, either from:
\begin{itemize}
    \item[(a)] the unit-mean discrete distribution given by 
    $$
    \begin{cases}
        (1-p) & \text{ with probability } 1/3,\\
        1 & \text{ with probability } 1/3,\\
        (1+p) & \text{ with probability } 1/3;
    \end{cases}
    $$
    \item[(b)] 
    a normal distribution with mean $1$ and variance $p$, i.e., $\mathcal{N}(1,p)$.
\end{itemize}
We repeat this sampling procedure, varying the parameter $p$ from $0$ (modeling perfect forecasts) to $p=0.3$ (modeling very noisy forecasts). Using these perturbed time series, we apply the Linear and Step heuristics and calculate the improvement over the static policy using~\eqref{Deltap2Lin}. Figure \ref{fig:perturbations} shows that until the magnitude $p$ of the perturbations is below $25-30\%$ our proposed dynamic dimensioning heuristics still outperform the static approach. 

\section{Discussion} \label{Discussion}
In this paper we took the Nordic grid as a case study and modeled the underlying frequency dynamics as a generalized Ornstein–Uhlenbeck (diffusion) process with non-constant drift. We used it to assess the reliability level of the grid, given the current volatility and available (static) FCR-N and FCR-D policies. We proposed and assessed multiple heuristic dynamic FCR-N policies and compared them with the current static one, and showed that, in general, a dynamic dimensioning approach can offer more efficient and more effective allocation of primary reserves.

Instead of applying heuristics to determine the $r_N^i$ (the amount of FCR-N volume reserved at hour $i$) for each hour of the year, we also considered identifying it as the solution of an optimization problem, where the objective could, for instance, be to minimize the sum of exceedance probabilities, i.e., $\min_{r_N^i}  \sum_{i=1}^{8760} p_2\left(r_N^i\right)$, such that the total annual budget is not exceeded (i.e., $\sum_{i=1}^{8760} r_N^i  \leq 5.256.000$) and the exceedance probability for each individual hour remains small (i.e., $2 p_2 (r_N^i) \leq 2\%$, for every $i=1,\ldots,8760$). However, in all cases considered, the resulting optimization problem is non-linear and we could not prove it to be convex, making it hard to solve, while heuristics already convey the message we want to get across. \new{The practical contribution of this work lies in demonstrating the potential of dynamic dimensioning for FCR. While our focus is on relatively simple heuristics, TSOs could readily adopt more sophisticated approaches, such as machine learning techniques, to further enhance the benefits of dynamic dimensioning for FCR. 

On the theoretical side, our contribution concerns the modeling of the underlying frequency dynamics. Nevertheless, alternative mathematical models for frequency deviations leading to different formulations of $p_1$, $p_2$, and $p_3$ could also be combined with the Linear and Step Heuristics to illustrate the advantages of dynamic dimensioning in FCR.}

Our findings are also relevant for larger synchronous areas with traditionally higher inertia, such as Continental Europe. As renewable penetration increases and inertia declines also in those areas, dynamic dimensioning of primary reserves such as FCR may become increasingly important. The diffusion model we introduced for the Nordic synchronous area can also be naturally adapted to reflect the specific dynamics of the ancillary services in other regions, as long as they are also linearly activated as a function of the frequency deviation.

\section{Future research} \label{Outlook}
In our study, a key assumption is the ability to reliably forecast the standard deviation $\sigma$ of system frequency on an hourly basis prior to the activation of FCR-N and FCR-D. While previous efforts have addressed the prediction of hourly imbalances \cite{PoCBelgium}, \cite{NSide}, further research should examine the sensitivity of our results to forecast errors beyond the basic analysis presented here. A potential approach would be to integrate our diffusion process approach into a physics-informed machine learning setting, as done in \cite{Kruse}. 
Nonetheless, the `StepFloor' heuristic that we propose in \eqref{StepFloor} is very conservative in the sense that it does not take additional risk based on forecasts of $\sigma$ and still illustrates the effectiveness of dynamic dimensioning.

Another avenue for future research could be to extend our work to include the effects of prices and markets, especially in the interplay between FCR-N and FCR-D.  In this context, the diffusion model could be extended from one to two dimensions by treating $x_D$ as a state variable rather than a fixed parameter. A key challenge in this extension would be establishing the convexity of the resulting optimization problem.

Lastly, our diffusion model currently assumes Gaussian noise to capture the stochasticity. While this provides a reasonable approximation for the bulk of the frequency distribution, a more accurate characterization of the tails, potentially heavier than Gaussian, is likely required for FCR-D applications \cite{Kraljic}. \new{We remark that, even though we assume Gaussian noise in our model, the sensitivity analysis on the forecasting error presented in Subsection \ref{sensitivity} is also relevant to gauge the potential impact of this assumption. An extension of our model could incorporate a jump component in the noise, as proposed in a different context by \cite{PatchZwart}. Such jump-diffusion models sometimes remain computationally feasible, particularly when using mixtures of exponential distributions for jumps, and merit exploration in future work.}

\section*{Acknowledgments}
The authors thank Daniel Br\o{}nsted Vesterlund Nielsen, Marco Saretta, Johanna Vorwerk, 
Camilla Damian, 
Thomas Dalgas Fechtenburg, Pia Ruokolainen, Mikko Kuivaniemi, 
and Robert Eriksson 
for useful discussions.

{\appendices
\section{Proof of Theorem \ref{Thm: StatDistr}}\label{App: Proof}

\begin{proof}
The outline of the proof is as follows. 
First, we calculate the component densities $\tau_j$ using the theory in \cite{Klebaner} and the symmetry of the diffusion process. Next, we use the theory in \cite{BrowneWhitt} to combine the component densities into one overall density.

\subsection{Calculating component densities}
In~\cite[Section 6.9]{Klebaner}, it is stated that under appropriate conditions on the drift coefficient $\mu(\cdot)$ and diffusion coefficient $\sigma(\cdot)$, a stationary distribution with density $\tau(\cdot)$ exists and can be found by calculating an explicit formula given by formula (6.69) on page 171 therein. For the diffusion process~\eqref{Diffproc}, it holds that $\mu(\cdot)=-\alpha(\cdot)$ with $\alpha(\cdot)$ as defined in \eqref{Alpha}, $\sigma(\cdot) = \sigma$ is a constant, and it can be shown that the required conditions hold so that the resulting equivalent of the above-mentioned formula (6.69) in our case comes down to calculating:

\begin{equation}
\label{pi}
    \tau(f) = \frac{C}{\sigma^2} \, \text{exp}\left( \, \int_{f_0}^{f} \frac{-2 \, \alpha(q) \, q}{\sigma^2} \, dq\right),
\end{equation}
for some $f_0 \in \mathbb{R}$ and where $C$ is found from $\int \tau(f) df = 1$ and for this purpose $\alpha(\cdot):= \alpha(r_N;x_D, \cdot)$ for given $r_N$ and $x_D$.
We now apply \eqref{pi} to the various segments as defined in \eqref{segments} to obtain the component densities $\tau_j(\cdot), \quad j=1,...,6$ and then combine them in an appropriate way to obtain the general stationary density $\tau(\cdot)$. 

Starting with $\tau_6(\cdot)$, formula \eqref{pi} translates to
\begin{align}
    \tau_6(f) \, &\stackrel{\eqref{Alpha}}{=} \frac{C_6}{\sigma^2} \, \text{exp}\left( \, \int_{0.5}^{f} \frac{-2 \, (r_N+x_D) \, q}{\sigma^2} \, dq\right) \nonumber \\
    &= \frac{C_6}{\sigma^2} \, \text{exp}\left(\frac{r_N+x_D}{4\sigma^2}\right) \text{exp}\left(-f^2\frac{r_N+x_D}{\sigma^2}\right),  \label{pi6}
\end{align}
for $f \in \text{I}_6= (0.5,\infty)$.
Since $\int_{0.5}^{\infty} \tau_6(f) df = 1$, 
\begin{align}
    C_6 &= \frac{\sigma^2 \, \text{exp}\left(-\frac{r_N+x_D}{4\sigma^2}\right)} {\int_{0.5}^{\infty} \text{exp}\left(-f^2\frac{r_N+x_D}{\sigma^2}  \right)df} \label{C6}.
\end{align}
Plugging this back in \eqref{pi6}, gives
\begin{align}
    \tau_6(f) \, 
    &= \frac{\text{exp}\left(-f^2\frac{r_N+x_D}{\sigma^2}\right)}{K_6(r_N;x_D, \sigma)}, \hspace{2mm} f\in \text{I}_6, \label{pi6def}
\end{align} 
with $K_6(r_N;x_D, \sigma):=\int_{0.5}^{\infty} \text{exp}\left(-y^2\frac{r_N+x_D}{\sigma^2}  \right)dy$.
Using symmetry, 
%
\begin{align}
    \tau_1(f) \, &= \tau_6(-f) \, = \tau_6(f), 
\end{align}
for $f \in \text{I}_1$ 
and
%
 $K_1(r_N;x_D, \sigma)
 = K_6(r_N;x_D, \sigma).$ 

Similarly, one can derive the expressions
for $\tau_2(\cdot), \tau_3(\cdot), \tau_4(\cdot)$ and $\tau_5(\cdot)$ as in \eqref{taui}.

\subsection{Calculating overall density}
The expression for all component densities $\tau_j(\cdot)$ for $j=1, ..., 6$, can be combined to get to the overall stationary density $\tau(\cdot)$, using the methodology described in \cite{BrowneWhitt} in particular, formulas (18.3)--(18.6) therein. 
The stationary density for a piecewise-continuous diffusion process is for $ -\infty = s_0 < s_1 < s_2 < s_3 < s_4 < s_5 < s_6 = \infty$ of the form 
    $\tau(f) = p_j \tau_j(f)$, 
with $s_{j-1} < f \leq s_j$, for $ j = 1, ..., 6$ and
\begin{align}
&\int_{s_{j-1}}^{s_j} \tau_j(f) \, df = 1, \hspace{2mm}
\sum_{j=1}^6 p_j = 1, \hspace{2mm} p_j := \frac{\rho_j}{\sum_{k=1}^6 \rho_k}, \label{pj}\\
&\rho_1 = 1, \text{ and } \rho_j := \prod_{k=2}^j \frac{\sigma_{k-1}^2 \tau_{k-1}(s_{k-1}-)}{\sigma_k^2 \tau_k(s_{k-1}+)}, \, 2 \leq j \leq 6.\label{rhos}
\end{align}
By choosing $s_1 := -0.5, s_2 := -0.1, s_3 := 0.0, s_4 := 0.1, s_5 := 0.5$, we recover the intervals
$\text{I}_j= (s_{i-1}, s_i]$, $ j = 1, ..., 6$.
Since in our case $\sigma_j = \sigma$ for all $j=1, \ldots, 6$, and the component $\tau_i$'s are all continuous, computing the $\rho_i$'s comes down to (writing $K_i:=K_i(r_N;x_D,\sigma)$ and using \eqref{rhos}) 
\begin{align*}
    \rho_1 &= 1 \text{ (by definition)} \\
    \rho_2  
    &= \frac{\tau_1(-0.5)}{\tau_2(-0.5)}  = \frac{\frac{1}{K_1}\text{exp}\left(-(-0.5)^2\frac{r_N+x_D}{\sigma^2}\right)}{\frac{1}{K_2}\text{exp}\left(-(-0.5)^2\frac{4r_N-x_D}{4\sigma^2}+(-0.5)^3\frac{5x_D}{3\sigma^2}\right)} \\
    &= \frac{K_2}{K_1} \text{exp}\left(-\frac{5x_D}{48\sigma^2}\right),
\end{align*}
and similarly, 
\begin{align*}
    \rho_3  = \rho_4 = &\frac{K_3}{K_1}\text{exp}\left(-\frac{31x_D+r_N}{300\sigma^2}\right),\\
    \rho_5  = &\frac{K_2}{K_1} \text{exp}\left(-\frac{5x_D}{48\sigma^2}\right).
\end{align*}
Finally, we write out the intermediate steps for $\rho_6$ to illustrate how the symmetry
\begin{equation}
\tau_1(f) = \tau_6(-f), \, \tau_2(f) = \tau_5(-f) \text{ and } \tau_3(f) = \tau_4(-f), \label{tausymmetry}
\end{equation}
is used in the calculation of the various $\rho_j$ above:
\begin{align*}
    \rho_6 
    \quad & = \frac{\tau_1(-0.5)}{\tau_2(-0.5)} \frac{\tau_2(-0.1)}{\tau_3(-0.1)} \frac{\tau_3(0.0)}{\tau_4(0.0)} \frac{\tau_4(0.1)}{\tau_5(0.1)} \frac{\tau_5(0.5)}{\tau_6(0.5)} \nonumber \\
    \quad & \stackrel{\eqref{tausymmetry}}{=} \frac{\tau_6(0.5)}{\tau_6(0.5)} \frac{\tau_5(0.1)}{\tau_5(0.1)} \frac{\tau_4(0.0)}{\tau_4(0.0)} \frac{\tau_4(0.1)}{\tau_4(0.1)} \frac{\tau_5(0.5)}{\tau_5(0.5)} = 1.
\end{align*}
Note that we have $\rho_1  = \rho_6, \, \rho_2 = \rho_5$, and $\rho_3 = \rho_4$, as we would expect because of the symmetry of the diffusion. Hence, using \eqref{pj}, we find that $p_1$ and $p_6$ are equal to
\begin{equation*}
 \frac{\rho_6}{\sum_{k=1}^6 \rho_k}
= \frac{1}{2 + 2 \frac{K_2}{K_1} \text{exp}\left(-\frac{5x_D}{48\sigma^2}\right) + 2\frac{K_3}{K_1}\text{exp}\left(-\frac{31x_D+r_N}{300\sigma^2}\right)},     
\end{equation*}
and similarly,
\begin{align}
p_2 = p_5 
&= \frac{\frac{K_2}{K_1} \text{exp}\left(-\frac{5x_D}{48\sigma^2}\right)}{2 + 2 \frac{K_2}{K_1} \text{exp}\left(-\frac{5x_D}{48\sigma^2}\right) + 2\frac{K_3}{K_1}\text{exp}\left(-\frac{31x_D+r_N}{300\sigma^2}\right)}, \nonumber \\
p_3 = p_4 
&= \frac{\frac{K_3}{K_1}\text{exp}\left(-\frac{31x_D+r_N}{300\sigma^2}\right)}{2 + 2 \frac{K_2}{K_1} \text{exp}\left(-\frac{5x_D}{48\sigma^2}\right) + 2\frac{K_3}{K_1}\text{exp}\left(-\frac{31x_D+r_N}{300\sigma^2}\right)}, \nonumber
\end{align}
from which it can be easily seen that $\sum_{i=j}^6p_j=1$ and indeed these $p_j$'s equal those in \eqref{probab}. Using indicator functions, we then obtain~\eqref{tau}.
\end{proof}

 
%

\bibliographystyle{IEEEtran}
\bibliography{finalrevision}
\end{document}